\makeatletter
\let\@twosidetrue\@twosidefalse
\let\@mparswitchtrue\@mparswitchfalse
\makeatother
\documentclass{llncs}
\usepackage[paperheight=235mm, paperwidth=155mm,textwidth=12.2cm,textheight=19.3cm]{geometry}
\usepackage[T1]{fontenc}

\makeatletter
\RequirePackage[bookmarks,unicode,colorlinks=true,breaklinks=true]{hyperref}%
   \def\@citecolor{blue}%
   \def\@urlcolor{blue}%
   \def\@linkcolor{blue}%

\def\orcidID#1{\smash{\href{http://orcid.org/#1}{\protect\raisebox{-1.25pt}{\protect\includegraphics{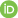}}}}}
\makeatother

\usepackage{graphicx} 
\usepackage{caption}
\usepackage{subcaption}
\usepackage{booktabs}
\usepackage[dvipsnames,table]{xcolor}
\usepackage{xspace}
\usepackage{array}
\usepackage[inline,shortlabels]{enumitem}
\usepackage{wrapfig}
\usepackage{amsmath,amssymb,ifthen}
\usepackage{stmaryrd}
\usepackage{setspace}
\usepackage{cite}

\graphicspath{{img/},{figures/}}

\newcommand{\modest}{\textsc{\mbox{Modest}}\xspace}
\newcommand{\toolset}{\text{\mbox{Modest} Toolset}\xspace}

\newcommand{\tool}[1]{\textsf{#1}\xspace}
\newcommand{\lang}[1]{\textsc{#1}\xspace}
\newcommand{\modes}{\tool{modes}}
\newcommand{\mcsta}{\tool{mcsta}}
\newcommand{\dtcontrol}{\tool{dtControl}}

\newcommand{\eg}{e.g.\ }
\newcommand{\ie}{i.e.\ }
\newcommand{\ZZ}{\ensuremath{\mathbb{Z}}\xspace}
\newcommand{\QQ}{\ensuremath{\mathbb{Q}}\xspace}

\newcommand{\nrepisodes}{\ensuremath{N}}
\newcommand{\tuple}[1]{\langle{#1}\rangle}
\newcommand{\St}{\ensuremath{S}}
\newcommand{\Act}{\ensuremath{A}}

\newcommand{\FO}{\textsf{\itshape F}\xspace}  
\newcommand{\PO}{\textsf{\itshape P}\xspace}  
\newcommand{\powerset}[1]{\ensuremath{2^{#1}}\xspace}
\newcommand{\Dist}[1]{\ensuremath{\mathit{Dist}({#1})}\xspace}
\newcommand{\defeq}{\mathrel{\vbox{\offinterlineskip\ialign{\hfil##\hfil\cr{\tiny \rm def}\cr\noalign{\kern0.30ex}$=$\cr}}}}
\newcommand{\support}[1]{\ensuremath{\mathit{spt}({#1})}\xspace}

\newcommand{\xdasharrow}[2][->]{
\tikz[baseline=-\the\dimexpr\fontdimen22\textfont2\relax]{
\node[anchor=south,font=\scriptsize, inner ysep=1.5pt,outer xsep=2.2pt](x){#2};
\draw[shorten <=3.4pt,shorten >=3.4pt,dashed,#1](x.south west)--(x.south east);
}
}\newcommand{\xtrp}[1]{\xrightarrow{\protect{\raisebox{-1.75pt}[0pt][0pt]{\ensuremath{\scriptstyle{#1}}}}}}
\newcommand{\xtrm}[1]{\xdasharrow{\protect{\raisebox{-0.5pt}[0pt][0pt]{\ensuremath{{#1}}}}}}

\newcounter{codelinenr}
\newcommand{\lcnt}{\refstepcounter{codelinenr}\makebox[2ex][r]{{\textrm{\bf\relsize{-3}\thecodelinenr}}}}
\newcommand{\resetcodelinenr}{\setcounter{codelinenr}{0}}

\usepackage{listings}
\newlength{\MaxSizeOfLineNumbers}%
\makeatletter
\lst@AddToHook{OnEmptyLine}{\vspace{-0.75\baselineskip}}
\lst@Key{countblanklines}{true}[t]%
    {\lstKV@SetIf{#1}\lst@ifcountblanklines}
\lst@AddToHook{OnEmptyLine}{%
    \lst@ifnumberblanklines\else%
       \lst@ifcountblanklines\else%
         \advance\c@lstnumber-\@ne\relax%
       \fi%
    \fi}
\makeatother
\usepackage[vlined,linesnumbered]{algorithm2e}

\SetAlCapFnt{\footnotesize}\SetAlCapNameFnt{\footnotesize}
\SetAlCapSkip{2ex plus 0.5ex minus 1ex}
\SetAlgoCaptionSeparator{.}
\SetEndCharOfAlgoLine{}
\SetArgSty{textrm}
\SetKwInOut{Input}{Input}\SetKwInOut{Output}{Output}

\SetCommentSty{mycommfont}
\SetKw{Break}{break}
\SetKwProg{Type}{type}{}{end}
\SetKwProg{Function}{function}{}{end}
\SetKwFor{ForEach}{foreach}{do}{}
\SetKwFor{WhileTrue}{repeat}{}{end}
\SetKw{Continue}{continue}
\SetKwRepeat{DoWhile}{repeat}{while}
\SetKwInput{KwReq}{Requires}

\usepackage{xspace}
\newcommand{\set}[1]{\ensuremath{\{\,#1\,\}}}

\SetAlFnt{\footnotesize}
\SetVlineSkip{0pt}
\SetKwProg{Fn}{}{}{}\SetKwFunction{Episode}{\textcolor{red}{\optima Episode}}%

\let\oldnl\nl
\newcommand{\nonl}{\renewcommand{\nl}{\let\nl\oldnl}}

\usepackage{cleveref}

\Crefname{figure}{Fig.}{Figs.}
\crefname{figure}{fig.}{figs.}
\Crefname{tabular}{Tab.}{Tabs.}
\crefname{tabular}{tab.}{tabs.}
\Crefname{section}{Sect.}{Sects.}
\crefname{section}{sect.}{sects.}
\Crefname{equation}{Eq.}{Eqs.}
\crefname{equation}{eq.}{eqs.}
\creflabelformat{equation}{#2#1#3}
\Crefname{definition}{Def.}{Defs.}
\crefname{definition}{def.}{defs.}
\crefname{lstlisting}{listing}{listings}
\Crefname{lstlisting}{Listing}{Listings}
\crefname{algorithm}{alg.}{algs.}
\Crefname{algorithm}{Alg.}{Algs.}

\usepackage{pgfplots}
\usepackage{tikz}
\usetikzlibrary{arrows,calc,automata,positioning,decorations.pathreplacing,shapes.geometric}
\tikzset{align at top/.style={baseline=(current bounding box.north)}}
\tikzstyle{every node}=[font=\scriptsize]
\tikzstyle{server} = [draw,fill=white,ellipse,thick,align=center,inner sep=0pt,minimum size=4.5mm]
\tikzstyle{waiting} = [draw,fill=white,rectangle,thick,align=center,inner sep=4pt, minimum size=4.5mm]
\tikzstyle{state} = [draw,fill=white,circle,thick,align=center,inner sep=0pt,minimum size=4.5mm]
\tikzstyle{lstate} = [draw,fill=white,rectangle,rounded corners,thick,align=center,inner sep=2pt,minimum size=4.5mm]
\tikzstyle{dot} = [fill,circle,inner sep=0mm,minimum size=1.25mm,line width=0mm]

\newcolumntype{L}[1]{>{\raggedright\let\newline\\\arraybackslash}p{#1}}
\newcolumntype{C}[1]{>{\centering\let\newline\\\arraybackslash}p{#1}}
\newcolumntype{R}[1]{>{\raggedleft\let\newline\\\arraybackslash}p{#1}}

\definecolor{lightblue}{RGB}{231,255,255}
\definecolor{lightred}{RGB}{255,224,224}
\definecolor{lightorange}{RGB}{255,239,223}
\definecolor{lightyellow}{RGB}{255,255,224}
\definecolor{lightpurple}{RGB}{255,224,255}
\definecolor{darkerred}{RGB}{64,0,0}
\definecolor{darkred}{RGB}{128,0,0}
\definecolor{darkblue}{RGB}{0,0,128}
\definecolor{darkorange}{RGB}{128,64,0}
\definecolor{darkpurple}{RGB}{128,0,128}

\makeatletter
\def\THICKhrulefill{\leavevmode \leaders \hrule height 5pt\hfill \kern \z@}
\makeatother

\makeatletter%
\g@addto@macro\normalsize{%
  \setlength\abovedisplayskip{3pt}%
  \setlength\belowdisplayskip{3pt}%
  \setlength\abovedisplayshortskip{-3pt}%
  \setlength\belowdisplayshortskip{3pt}%
}%
\makeatother

\title{Digging for Decision Trees: A Case Study\\in Strategy Sampling and Learning}

\author{%
Carlos E. Budde\inst{1}\,\orcidID{0000-0001-8807-1548}
\and Pedro R. D'Argenio\inst{2,3}\,\orcidID{0000-0002-8528-9215}
\and Arnd Hartmanns\inst{4}\,\orcidID{0000-0003-3268-8674}
}
\institute{
University of Trento, Trento, Italy
\and Universidad Nacional de Córdoba, Córdoba, Argentina
\and CONICET, Córdoba, Argentina
\and University of Twente, Enschede, The Netherlands
}

\begin{document}

\maketitle

\begin{abstract}
We introduce a formal model of transportation in an open-pit mine for the purpose of optimising the mine's operations.
The model is a network of Markov automata (MA); the optimisation goal corresponds to maximising a time-bounded expected reward property.
Today's model checking algorithms exacerbate the state space explosion problem by applying a discretisation approach to such properties on MA.
We show that model checking is infeasible even for small mine instances.
Instead, we propose statistical model checking with lightweight strategy sampling or table-based Q-learning over untimed strategies as an alternative to approach the optimisation task, using the Modest Toolset's \modes tool.
We add support for partial observability to \modes so that strategies can be based on carefully selected model features, and we implement a connection from \modes to the \dtcontrol tool to convert sampled or learned strategies into decision trees.
We experimentally evaluate the adequacy of our new tooling on the open-pit mine case study.
Our experiments demonstrate the limitations of Q-learning, the impact of feature selection, and the usefulness of decision trees as an explainable representation.
\end{abstract}

\section{Introduction}

The model of Markov decision processes (MDPs)~\cite{Bel57,Put94} precisely captures the interplay of controllable or uncontrollable (nondeterministic) choices with randomness.
It is simple yet versatile, which has made it popular in finance and operations research~\cite{BR11}, in machine learning as the conceptual foundation of reinforcement learning~\cite{SB98}, and in verification as the core formalism of probabilistic model checking (PMC)~\cite{BAFK18,HJQW23}.
MDPs are fully discrete:
they transition between discrete states in discrete time; the number of choices per state is countable or finite; and the outcome of a choice is sampled from a discrete probability distribution.
Modelling some applications, however, requires an explicit representation of continuous real time.
Examples include performance evaluation scenarios such as computing the expected number of requests handled by a server per second, or determining the probability for the execution delay of a real-time task to exceed a certain bound.
Models based on timed automata (TA)~\cite{AD94} offer a notion of ``hard'' real time with fixed deterministic delays or nondeterministic intervals.
The tool \tool{Uppaal}~\cite{BDL04} prominently supports the analysis of TA models via traditional and and statistical model checking (SMC)~\cite{BDLMPLW12}.
When events occur at random times with a known rate, a ``soft'' real-time model with stochastic, exponentially-distributed delays based on continuous-time Markov chains (CTMCs) is more appropriate.

In this paper, we present a novel case study about the optimisation of transport operations in an open-pit mine.
It requires consideration of both controllable choices and stochastic time.
The optimisation goal is to schedule trucks carrying material from shovels to dumps so that the expected total amount of material carried at the end of a shift is maximised.
In abstract terms, this scenario is naturally modelled as a network of Markov automata (MA)~\cite{EHZ10,HH19}, which combine the features of MDPs and CTMCs in a compositional manner.
The optimisation goal can then be phrased as a query for the strategy maximising the value of a time-bounded expected accumulated reward property.

The analysis of MA is implemented by \tool{mcsta}~\cite{BHH21} and \tool{Storm}~\cite{HJKQV22} via PMC and by \modes via SMC~\cite{BDHS20}.
We use \tool{mcsta} and \modes in this paper; they are part of the \toolset~\cite{HH14}, available online at \href{https://www.modestchecker.net/}{modestchecker.net}.
Whereas PMC~\cite{BAFK18} uses iterative numeric algorithms on an in-memory representation of the entire MA's state space (or of a subset sufficient for an $\epsilon$-precise approximation~\cite{ABHK18}), SMC~\cite{AP18} performs a statistical evaluation of a large number of samples of the MA's behaviour.
Efficient PMC algorithms exist to compute the values of unbounded properties by analysing the MA's embedded MDP, and for time-bounded reachability probabilities~\cite{BF19,BHHK15}.
For time-bounded expected rewards, however, the only available PMC algorithm today uses a discretisation approach~\cite{Hat17}, which exacerbates the state space explosion problem of PMC by introducing many new states for the many discrete time steps needed for a reasonably precise result.
We show in \Cref{sec:ExperimentalResults} that even checking the embedded MDP of small variants of our case study is infeasible; analysing our property of interest using a discretisation-based approach is thus out of the question.

Our contributions are
(1)~the introduction of the new open-pit mining case study (\Cref{sec:CaseStudy}) and its MA model in the \modest modelling language~\cite{BDHK06,HHHK13} (\Cref{sec:ModellingAndAnalysis});
(2)~an experimental evaluation (\Cref{sec:ExperimentalResults}) of two approaches that extend SMC from estimation to optimisation (as required for MA) on several instances of varying sizes of the case study: smart lightweight strategy sampling (LSS)~\cite{DLST15,LST14} and (explicit table-based) Q-learning~\cite{SB98,WD92}, both implemented in \modes; and
(3)~two extensions to \modest and \modes to make LSS and Q-learning feasible and their outcomes explainable (\Cref{sec:ToolExtensions}).
Our first extension is support for partial observations of states by designating a subset of the model's variables as observable.
A good choice of observables reduces the sample space of strategies for LSS and the potential table size for Q-learning without removing much relevant information, making both methods more effective.
Their outcomes are an opaque strategy identifier for LSS and a large table of strategy choices for Q-learning, both of which are hardly useful for mine operators.
We thus created a new connection from \modes to \dtcontrol~\cite{AJKWWY21} to obtain a more explainable representation in the form of decision trees, implemented with a focus on minimal memory usage to be able to tackle large problems like the open-pit mining case.

\section{The Open-Pit Mining Case Study}
\label{sec:CaseStudy}

Material transportation is one of the most important aspects that
affects productivity in an open pit mine~\cite{AG02}.
Material needs to be transported from the extraction points to
different places depending on if the material contains ore or is
simply waste.
The material is loaded onto trucks by shovels that operate in a
region of a single type of material. 
In general, there could be different type of ores in
different regions, but we will consider only one type of ore here.  The
useful material containing ore is hauled by the trucks and dumped on
stockpiles that later will be taken to crushers at the beginning of the
ore processing line.  The
waste material is hauled and dumped in a separate designated
place.
Once a truck has dumped its load, it returns to \emph{some} shovel and
repeats the process.

Trucks may need to queue
and wait at some loading or dumping place until other trucks are loaded
or unloaded.
Therefore, an appropriate distribution of the trucks that minimises their waiting time (and any other possible non-productive time)
is crucial.
The problem of assigning trucks to shovels and dumping places during
production is called the \emph{truck dispatching
problem}~\cite{AG02,MA17}, which we address in this paper.
Using the \toolset, we propose a \emph{flexible truck
allocation}~\cite{MA17} approach to solve the problem.
In such an approach, trucks are assigned a next dumping place or shovel by a
dispatch system whenever they are done loading at a shovel or unloading at a dumping place, respectively.

\begin{figure}[t]
  \centering
  \includegraphics[width=.7\linewidth]{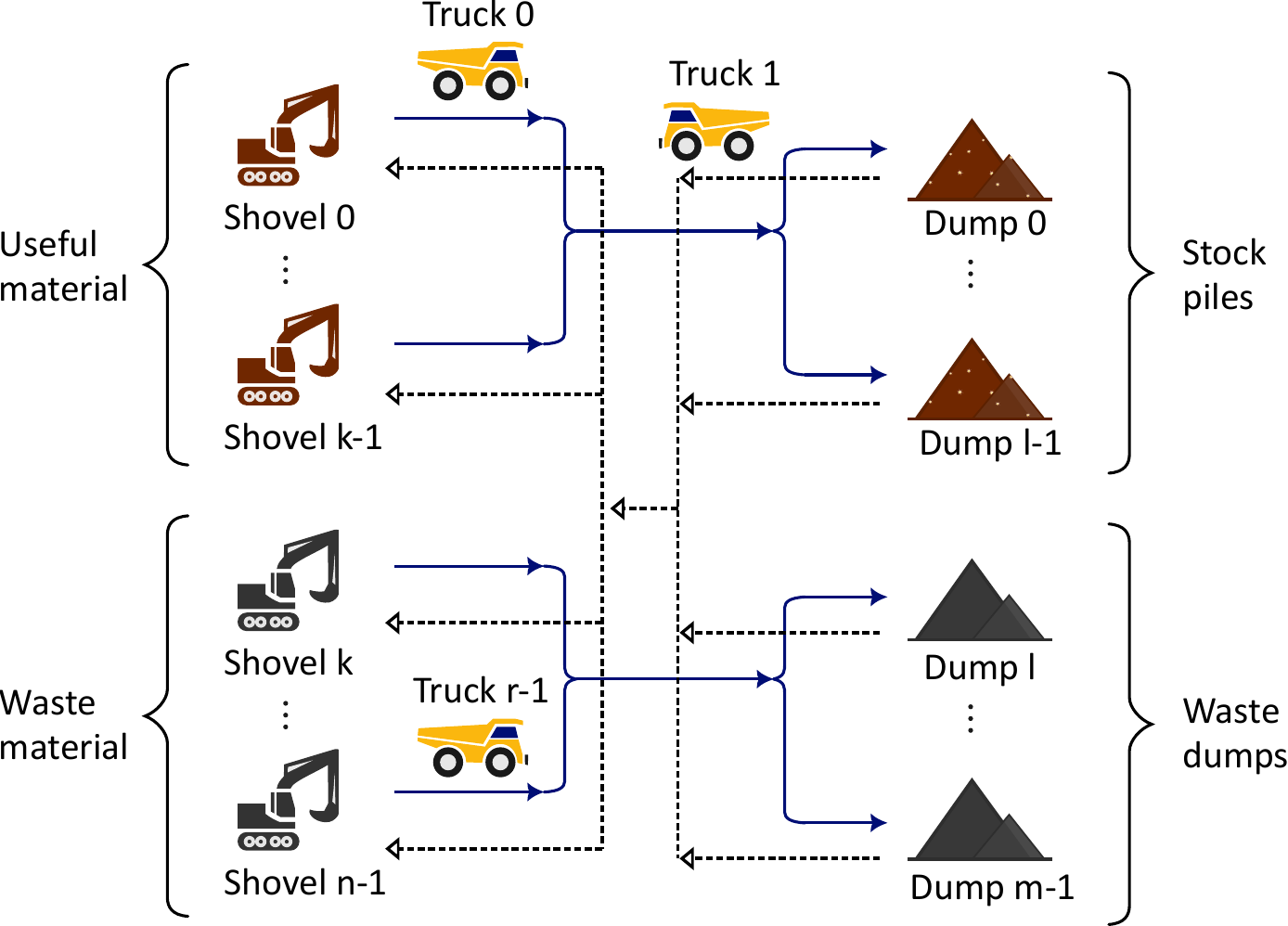}
  \caption{Schematic view of the open pit mine}\label{fig:schematic}
\end{figure}

\Cref{fig:schematic} shows a schematic view of the mine
operation in which $k$ shovels load useful material and $N-k$ shovels
load waste material.  There are $l$ dumping places associated to
stockpiles and $M-l$ waste dumps.  A truck loaded with
useful material can only be assigned a stockpile to haul the load to,
while a truck loaded with waste material can only be assigned a waste
dump.  Empty trucks that have just dumped
their load can be assigned to any shovel.
A truck takes time to move from one
point to the other.  Similarly, loading and dumping are also
activities that take time.
Our objective is to model this scenario with \modest and use the available tools of the \toolset in order to maximise the productivity of the trucks.
Concretely,
our optimisation goal is to maximise the total load of material transported in
one operation shift.

%
%
%

\section{Background: Modelling and Analysis}
\label{sec:Background}

In this section, we introduce all the existing concepts and technology necessary for our formal modelling and analysis of the open-pit mining case study.

\paragraph{Preliminaries.}
Given a set $S$, its powerset is $\powerset{S}$.
A probability distribution over $S$ is a function $\mu \colon S \to [0, 1]$ s.t.\ $\support{\mu} \defeq \set{ s \in S \mid \mu(s) > 0 }$ is countable and $\sum_{s \in \support{\mu}} \mu(s) = 1$.
$\Dist{S}$ is the set of all probability distributions over~$S$. 

\subsection{Markov Automata}

Markov automata combine MDPs and CTMCs in an orthogonal manner by providing two types of transitions: $s \xtrp{\;a\;}\mu$ as in MDP, and $s \xtrm{\;\lambda\;}s'$ as in  CTMC.
We now define Markov automata formally and describe their semantics.

\begin{definition}
\label{def:MA}
A \emph{Markov automaton} (MA) is a tuple
$M =\tuple{S, s_0, A, P, Q, \mathit{rr}, \mathit{br}}$
where
$S$ is a finite set of \emph{states} with \emph{initial state} $s_0 \in S$,
$A$ is a finite set of \emph{actions},
$P\colon \mathit{S} \to \powerset{A \times \Dist{S}}$ is the \emph{probabilistic transition} function,
$Q \colon \mathit{S} \to \powerset{\QQ \times S}$ is the \emph{Markovian transition} function,
$\mathit{rr} \colon S \to [0, \infty)$ is the \emph{rate reward} function, and
$\mathit{br} \colon S \times \mathit{Tr}(M) \times S \to [0, \infty)$ is the \emph{branch reward} function.
$\mathit{Tr}(M) \defeq \bigcup_{s \in S}{P(s) \cup Q(s)}$ is the set of all \emph{transitions}; it must be finite.
We require that $\mathit{br}(\tuple{s, \mathit{tr}, s'}) \neq 0$ implies $\mathit{tr} \in P(s) \cup Q(s)$.
\end{definition}
We also write $s \xtrp{a} \mu$ for $\tuple{a, \mu} \in P(s)$ and $s \xtrm{\lambda} s'$ for $\tuple{\lambda, s'} \in Q(s)$.
In $s \xtrm{\lambda} s'$, we call $\lambda$ the \emph{rate} of the Markovian transition.
We refer to every element of $\support{\mu}$ as a \emph{branch} of $s \xtrp{a}_P \mu$;
a Markovian transition has a single branch only. 
We define the \emph{exit rate} of $s \in S$ as $E(s) = \sum_{\tuple{\lambda, s'} \in Q(s)} \lambda$.

Intuitively, the semantics of an MA is that, in state $s$,
(1)~the probability to take Markovian transition $s \xtrm{\lambda} s'$ and move to state $s'$ within $t$ model time units is
${\lambda}/{E(s)} \cdot (1 - \mathrm{e}^{-E(s) \cdot t})$,
\ie the residence time in $s$ follows the exponential distribution with rate $E(s)$ and the choice of transition is probabilistic, weighted by the rates; and
(2)~at any point in time, a probabilistic transition $s \xtrp{a} \mu$ can be taken with the successor state being chosen according to~$\mu$.
We refer the interested reader to \eg\cite{Hat17} for a complete formal definition of this semantics.

\begin{figure}[t]
\begin{minipage}[b]{0.4\textwidth}
\centering
\begin{tikzpicture}[on grid,auto]
  \node[state] (s0) {I};
  \coordinate[above=0.3 of s0.north] (start);
  \node[] (me) [above left=0.36 and 1.1 of s0] {\small$M_e$:};
  \node[dot] (n0) [below left=0.5 and 0.4 of s0] {};
  \node[state] (s1) [below left=1.4 and 0.8 of s0] {$\text{S}_1$};
  \node[state] (s2) [below=1.4 of s0] {$\text{S}_2$};
  \node[state] (s3) [below right=1.4 and 0.8 of s0] {B};
  \node[state] (s4) [below=1.2 of s2] {X};
  ;
  \path[-]
    (s0) edge[bend right] node[left] {\texttt{a}} (n0)
  ;
  \path[->]
    (start) edge node {} (s0)
    (s0) edge[out=-20,in=90] node[right,pos=0.245] {\texttt{b}} (s3)
    (n0) edge[bend right] node[left] {$0.5$} (s1)
    (n0) edge[bend left] node[right] {$0.5$} (s2)
    (s1) edge[bend right=50,dashed] node[left] {$1$} (s4)
    (s2) edge[bend left=70,dashed] node[below,inner sep=1.5pt] {$\mathstrut2$} (s1)
    (s3) edge[bend left=50,dashed] node[right] {$4$} (s4)
    (s3) edge[bend left=70,dashed] node[below,inner sep=1.5pt] {$\mathstrut3$} (s2)
    (s4) edge[loop,dashed,looseness=5,out=-45,in=-90] node[right,pos=0.3,inner sep=1.5pt] {$1$} (s4)
  ;
\end{tikzpicture}
\caption{Example MA $M_e$}
\label{fig:ExampleMA}
\end{minipage}%
\begin{minipage}[b]{0.6\textwidth}
\centering
\begin{lstlisting}
action a, b;
int s;

alt {
:: a palt { :0.5: {= s = 2 =} :0.5: {= s = 1 =} }
:: b; alt {
   :: rate(3) {= s = 2 =}
   :: rate(4) {= s = 0 =}
   }
};
do {
:: when(s >= 1) rate(s) tau {= s-- =}
:: when(s == 0) rate(1) tau
}
\end{lstlisting}
\caption{MA $M_e$ in \modest}
\label{fig:ExampleMAModest}
\end{minipage}%
\end{figure}

\begin{example}
We show example MA $M_e$ without rewards in \Cref{fig:ExampleMA}.
Its initial state I has a choice between two probabilistic transitions with action labels \lstinline|a| and \lstinline|b|.
The former leads to each of states $\text{S}_1$ and $\text{S}_2$ with probability~$0.5$.
The dashed transitions are Markovian, labelled with their rates.
In state B, there is a race between two Markovian transitions; the expected time spent in B is $\frac{1}{7}$ time units.
\end{example}
An MA without Markovian transitions is an MDP; 
an MA without probabilistic transitions is a CTMC.
The separation of transitions into probabilistic and Markovian enables parallel composition with action synchronisation without the need to prescribe an ad-hoc operation for combining rates as would be necessary for CTMC or continuous-time MDP (CTMDP)~\cite{Put94}. 
For verification, after applying the semantics of parallel composition, we make the usual \emph{closed system} and \emph{maximal progress} assumptions: probabilistic transitions face no further interference and take place without delay.
The choice between multiple probabilistic transitions in a state remains nondeterministic, but all Markovian transitions can be removed from states that \emph{also} have a probabilistic transition. 

The behaviour of closed, deadlock-free MA $M$ is characterised by the set $\Pi$ of infinite timed paths in its semantics.
Let $\Pi_\mathit{fin}$ be the finite path prefixes.

\begin{definition}
\label{def:Strategy}
A \emph{strategy} is a function $\sigma \colon \Pi_\mathit{fin} \to \mathit{Tr}(M)$ s.t.\ $\forall s \in S\colon \sigma(s) = \mathit{tr}$ implies $\mathit{tr} \in P(s) \cup Q(s)$.
A \emph{time-dependent} strategy is in $S \times [0, \infty) \to \mathit{Tr}(M)$; a \emph{memoryless} strategy is in $S \to \mathit{Tr}(M)$.
\end{definition}
If we ``apply'' a strategy to an MA, it removes all nondeterminism, and we are left with a stochastic process whose paths can be measured and assigned probabilities according to the rates and distributions in the (remaining) MA.
We again refer the interested reader to \eg\cite{Hat17} for a fully formal definition.

Given an MA model, we are interested in determining the maximum (supremum) or minimum (infimum) value over all strategies of the following properties:
\begin{description}
\item[Expected accumulated reachability rewards:]
Compute the expected value of the random variable that assigns to $\pi$ the sum of its branch rewards and of each state's rate reward multiplied with the residence time in that state, up to the first state in goal set $G \subseteq S$.
This is well-defined if the maximum (minimum) probability to reach $G$ is $1$; otherwise, we define the minimum (maximum) expected value to be~$\infty$.
Memoryless strategies suffice to achieve optimal results (\ie the maximum and minimum expected values).
\item[Time-bounded expected accumulated rewards:]
Compute the expectation as above, but instead of stopping at goal states, stop once a given amount of time $\mathcal{T} \in [0, \infty)$ has elapsed.
Time-dependent strategies suffice.
\end{description}
Expected reachability rewards can be computed via standard MDP model checking by considering the MA's embedded MDP, \ie replacing all remaining Markovian transitions out of a state by a single probabilistic transition, using the rates as weights to determine the branch probabilities.
For time-bounded expected rewards, the best choice among probabilistic transitions may depend on the time remaining to $\mathcal{T}$---thus the need for time-dependent strategies.
The only available model checking algorithm for this type of property discretises the remaining time~\cite{Hat17}, exacerbating the state space explosion problem.
We therefore propose to use SMC instead, which however cannot directly deal with the optimisation problem induced by the nondeterministic choices.
We describe the two methods to deal with finding (near-)optimal strategies in SMC in \Cref{sec:SMC} below.

\subsection{Modest for Markov Automata}

\modest~\cite{BDHK06,HHHK13} is the \underline{mo}delling and \underline{de}scription language for \underline{s}tochastic \underline{t}imed systems.
It provides process-algebraic operations such as parallel and sequential composition, parameterised process definitions, process calls, and guards to construct complex models from small reusable parts.
Its syntax leans on commonly used programming languages, and it provides conveniences such as loops and an exception handling mechanism.
To specify complex behaviour in a succinct manner, \modest provides variables of standard basic types (\eg \texttt{bool}, \texttt{int}, or bounded \texttt{int}), arrays, and user-defined recursive datatypes akin to functional programming languages.
We introduce \modest for modelling MA by example:

\begin{example}
\Cref{fig:ExampleMAModest} shows a \modest model whose concrete semantics is our example MA~$M_e$.
Choices between multiple transitions are implemented with the \lstinline|alt| construct.
Probabilistic transitions can be labelled with user-defined action names like \lstinline|a| and \lstinline|b|, while Markovian transitions---for which a \lstinline|rate| is given---must use the predefined non-synchronising action \lstinline|tau|.
The \lstinline|palt| construct implements the probability distributions of probabilistic transitions; \lstinline|;| is the sequential composition operator.
Assignments \lstinline|s = 2| are given in assignment blocks \lstinline|{= ... =}|; if a block has multiple assignments, they are executed atomically.
\end{example}

\subsection{Analysis of Markov Automata via Statistical Model Checking}
\label{sec:SMC}

Monte Carlo methods such as \emph{statistical model checking} (SMC)~\cite{AP18}, which is in essence discrete-event simulation~\cite{Law15} for formal models and properties, can estimate expected rewards in CTMCs.
To do so, the SMC algorithm \mbox{(pseudo-)}\allowbreak{}randomly samples $n$ finite paths---\emph{simulation runs}---through the CTMC, collects each path's accumulated reward value, and returns the average of the collected values.
The result is correct up to a statistical error and confidence depending on $n$.
We assume that we can effectively perform simulation runs on a high-level
description of the MA (\eg in \modest).
Then, in contrast to PMC, SMC does not need to store the CTMC's entire state space and thus runs in constant memory.
SMC is easy to parallelise and distribute on multi-core systems and compute clusters.

The simulation of nondeterministic models like MDP or MA, however, requires a strategy to resolve the nondeterminism during simulation.
Ideally, such a strategy should be the optimal one, \ie one that results in the maximum or minimum expected reward value.
Obtaining optimal strategies experimentally is infeasible in any sensible model.
Instead, best-effort methods to find near-optimal strategies have been devised, notably based on \emph{strategy sampling} and~\emph{learning}.

\subsubsection{Lightweight strategy sampling}

(LSS) was devised for MDP~\cite{LST14}.
On MDP~$M\!$,
\begin{enumerate}[(i)]
\item%
  it randomly selects a set $\varSigma$ of $N$ strategies, each identified by a fixed-size integer (\eg of 32 bits as in our   implementation),
\item\label{enum:LSS:Step2}%
  employs a heuristic (that involves simulating the DTMCs $M|_\sigma$ resulting from applying a strategy $\sigma\in\varSigma$ to $M$) to select the $\sigma_\mathit{\!opt}\in\varSigma$, $\mathit{opt} \in \set{\mathrm{max}, \mathrm{min}}$, that appears to induce the highest/lowest probability, and finally
\item\label{enum:LSS:Step3}%
  performs standard SMC on $M|_{\sigma_\mathit{\!opt}}$ to provide an estimate $\hat{R}_{\sigma_\mathit{\!opt}}$ for the optimal expected accumulated reward. 
\end{enumerate}
Unless $\varSigma$ happens to include an optimal strategy and the heuristic identifies it as such, $\hat{R}_{\sigma_\mathit{max}}$ is only an underapproximation (overapproximation) of the maximum (minimum) expected reward,
and subject to the statistical error of the final SMC step.
The effectiveness of LSS depends on the probability mass of the set of near-optimal strategies among the set of all strategies that we sample $\varSigma$ from:
It works well if a randomly selected strategy is somewhat likely to be near-optimal, but usually fails in cases where many decisions need to be made in exactly one right way in order to get any non-negligible reward at all.

\begin{algorithm}[t]
  \setstretch{1.1}
  \KwIn{MA $M$, iteration budget $K$, and strategy budget $N$ with $N\leq K$.}
  \KwOut{The maximising strategy $\sigma_{\max}$.}
  $\varSigma := \set{\sigma_i\mid\sigma_i = \text{sample uniformly from } \ZZ_{32}, 0\leq i\leq N }$\;\label{alg:lss:samplstrat}
  \While{$|\varSigma| > 1$}{
    \lForEach{$\sigma\in \varSigma$}{
      $\hat{R}_\sigma := \text{average of } \lceil{K/|\varSigma|}\rceil \text{ simulation run values for } \sigma$\label{alg:lss:sim}
    }
    $\varSigma := \set{\sigma\in \varSigma\mid \hat{R}_\sigma \text{ is among the } \lceil|\varSigma|/2\rceil \text{ highest values in } \set{ \hat{R}_\sigma' \mid \sigma' \in \varSigma }}$\label{alg:lss:discard}
  }
  $\sigma_{\max} := \text{the only remaining strategy identifier in } \varSigma$
  \caption{Lightweight strategy sampling with the smart sampling heuristics}\label{alg:lss}
\end{algorithm}

\paragraph{Smart sampling.}
We use an MA adaptation of LSS with the \emph{smart sampling} heuristics~\cite{DLST15} 
for step~\ref{enum:LSS:Step2}.
It is schematically presented in \Cref{alg:lss} for $\mathit{opt} = \mathrm{max}$.
It receives as inputs the MA, the strategy budget $N$, and the iteration budget~$K$.
$N$ determines how many strategies will be randomly selected while $K$ is the number of simulation runs to be performed in each iteration.
We require $N\leq K$ so that in the first round, we have at least one simulation per strategy.
After sampling the strategies (line~\ref{alg:lss:samplstrat}), the algorithm runs $\lceil{K/N}\rceil$ simulations for each strategy estimating accumulated rewards (line~\ref{alg:lss:sim}).
Afterwards, it discards the worst half of the strategies (line~\ref{alg:lss:discard}) and simulates the remaining ones with twice the number of simulation runs per strategy (line~\ref{alg:lss:sim} again).
The loop repeats until only one strategy remains, which is $\sigma_\mathit{max}$.
In this way, the number of simulation runs, and thus the runtime, grows only logarithmically in~$N$.

\paragraph{Lightweight strategies.}
The key to LSS is the constant-memory representation of strategies as (32-bit) integers.
It enables the algorithm to run in constant memory, which sets it apart from simulation-based machine learning techniques such as reinforcement learning, which need to store learned information (\eg Q-tables, see below) for each visited state.

\begin{algorithm}[t]
  \setstretch{1.1}
  \KwIn{MA $M$, time bound $\mathcal{T}$, strategy identifier $\sigma \in \ZZ_{32}$, hash function $\mathcal{H}$.}
  \KwOut{The reward $r$ accumulated along the sampled run.}
  $s := s_0$, $t := 0$, $r := 0$\tcp*{initialise current state, time, and reward}\label{alg:lss-sim:init}
  \While(\tcp*[f]{run until time bound is reached}){$t \leq \mathcal{T}$\label{alg:lss-sim:loop}}{
    \If(\tcp*[f]{$s$ has only Markovian transitions}){$P(s) = \emptyset$}{
      $\tuple{t',\tuple{\lambda,s'}} := \text{sample sojourn time and transition from } Q(s)$\label{alg:lss-sim:then1}\;
      $r := r + \min(t', \mathcal{T} \!- t) \cdot \mathit{rr}(s) + \mathit{br}(\tuple{s,\tuple{\lambda,s'},s'})$\tcp*{collect the reward}\label{alg:lss-sim:newrm}
      $t := t + t'$\tcp*{increase current time}\label{alg:lss-sim:then2}
    }
    \Else(\tcp*[f]{$s$ is a probabilistic state}\label{alg:lss-sim:else}){
      $\tuple{a, \mu} := (\mathcal{H}(\sigma.s) \mathbin\mathrm{mod} |P(s)| + 1)\text{-th element of }P(s)$\tcp*{select\:transition}\label{alg:lss-sim:else1}
      $s' := \text{sample the next state according to } \mu$\;\label{alg:lss-sim:else2}
      $r := r + \mathit{br}(\tuple{s,\tuple{a,\mu},s'})$\tcp*{collect the reward}\label{alg:lss-sim:newrp}
    }
    $s := s'$\tcp*{set new current state}\label{alg:lss-sim:news}
  }
  \caption{A single simulation run using a sampled strategy $\sigma$ in LSS}\label{alg:lss-sim}
\end{algorithm}

\Cref{alg:lss-sim}, the simulation called in line \ref{alg:lss:sim} of \Cref{alg:lss}, shows how this is done.
Apart from the MA $M$ and the time-bounded reward property's stopping time $\mathcal{T}$, its inputs include the strategy identifier $\sigma \in \ZZ_{32}$ and a (usually simple non-cryptographic) uniform deterministic hash function $\mathcal{H}$ that maps to values in $\ZZ_{32}$.
The algorithm is mostly self-explanatory.
Lines \ref{alg:lss-sim:then1}-\ref{alg:lss-sim:then2} take care of
the selection of the Markovian step, the time advance, and the update
of the reward for the Markovian case.
The notable part lies in the case of a choice between $k>1$ enabled probabilistic transitions (line~\ref{alg:lss-sim:else}).  Assuming some total order on the transitions, the $(\mathcal{H}(\sigma.s)\mathbin{\mathrm{mod}}k)$-th transition is selected, where $\sigma.s$ is the concatenation of the binary representations of $\sigma$ and $s$ (line~\ref{alg:lss-sim:else1}).
This selection procedure is deterministic, so we can reproduce the decision for state $s$ at any time knowing $\sigma$.
For nontrivial $\mathcal{H}$, it is also highly unpredictable: changing a single bit in $\sigma$ may result in a different decision for many states.
The selected transition is then used to sample the next state and the update of the reward (lines \ref{alg:lss-sim:else2} and~\ref{alg:lss-sim:newrp}).
Finally, after updating the accumulated reward and the current state, the loop continues until the time limit~$\mathcal{T}$ is reached.

Observe that \Cref{alg:lss-sim} implements a memoryless strategy, whereas time-de\-pen\-dent strategies would be needed to obtain optimal results for time-bounded reward properties.
This is a practical simplification to make LSS work effectively:
If we used time-dependent strategies directly by \eg feeding a floating-point representation of $\mathcal{T}$ into $\mathcal{H}$ as well, then all strategies would break down to behaving like the uniformly random strategy~\cite{DHS18};
if we used discretisation like in the model checking algorithm, the space of strategies would blow up so much that the probability of sampling a useful strategy would be negligible.

\subsubsection{Q-learning}

(QL)~\cite{WD92} is a popular method for reinforcement learning (RL)~\cite{SB98}, a machine learning approach to train agents to take actions maximising a reward in uncertain environments.
Mathematically, the agent in its environment can be described as \eg a MA\footnote{To ease the presentation, we assume actions to uniquely identify transitions per~state.}:
the agent chooses actions while the environment determines the probabilistic outcomes of the actions in terms of successor states.
QL maintains a Q-function $\mathcal{Q}\colon \St \times \Act \to [0, 1]$ stored in explicit form (as a so-called \emph{Q-table}) initialised to $0$
everywhere.
For MA, the Q-table only needs to store values for states with probabilistic transitions.

\begin{algorithm}[t]
  \setstretch{1.1}
  \KwIn{MA $M$, time bound $\mathcal{T}$, number of episodes $\nrepisodes$, family of learning rates $\{\alpha_i\}_{i=1}^\nrepisodes$, family of exploration likelihoods $\{\epsilon_i\}_{i=1}^\nrepisodes$, discount factor $\gamma$.}
  \KwReq{$s_0$ is a state with probabilistic transitions.}
  \KwOut{Table $\mathcal{Q}$ assigning expected discounted reward to state-action pairs.}
  \For{$i := 1$ \KwTo $\nrepisodes$}{
    $s := s_0$, $t := 0$\tcp*{initialise current state and time}
    \While(\tcp*[f]{run until time bound is reached}){$t \leq \mathcal{T}$}{
      $\tuple{a, \mu} := \text{sample uniformly from } P(s)$%
          \tcp*{$s$ has probabilistic transitions}\label{alg:q-learning:exex}
      \nonl
      $\phantom{\tuple{a, \mu} :=}
          \oplus_{\epsilon_i}
          \displaystyle
          \!\!\mathop{\textrm{arg max}}_{\tuple{a'\!,\mu'}\in P(s)}\!\! \mathcal{Q}(s,a')$\tcp*{$\oplus_{\epsilon_i}$: random choice with probability~$\epsilon_i$}
      $s' := \text{sample the next state according to } \mu$\;\label{alg:q-learning:xs}
      $r := \mathit{br}(\tuple{s,\tuple{a,\mu},s'})$\tcp*{collect the reward}\label{alg:q-learning:rew}
      $s := s'$, $s'' := s'$\tcp*{set new current state}
      \While(\tcp*[f]{while $s'$ has only Markovian transitions:}){$t \!\leq\! \mathcal{T} \wedge P(s') \!=\! \emptyset$\label{alg:q-learning:while}}{
        $\tuple{t',\tuple{\lambda,s''}} := \text{sample sojourn time and transition from } Q(s')$\;
        $r := r + \min(t', \mathcal{T} \!- t) \cdot \mathit{rr}(s') + \mathit{br}(\tuple{s',\tuple{\lambda,s''},s''})$\tcp*{collect reward}
        $s' := s''$, $t := t + t'$\tcp*{set new state and increase time}
      }\label{alg:q-learning:elihw}
      \vspace{3pt}$\mathcal{Q}(s,a) :=
           (1-\alpha_i) \cdot \mathcal{Q} (s,a)
           +
            \alpha_i \cdot \left(r + \gamma\cdot\max _{\tuple{a'',\mu''}\in P(s'')} \mathcal{Q}(s'',a'')\right)$\label{alg:q-learning:learn}
            \;
      $s := s''$\tcp*{set new current state}
    }
  }
  \caption{Q-learning algorithm for MA}
  \label{alg:q-learning}
\end{algorithm}

\paragraph{Algorithm.}
Using a family of \emph{learning rates}
$\{\alpha_i\}_{i=1}^\nrepisodes$, a family of \emph{exploration
likelihoods} $\{\epsilon_i\}_{i=1}^\nrepisodes$, and a discount
factor $\gamma \in (0, 1]$, $\nrepisodes$ learning \emph{episodes} are
performed following \Cref{alg:q-learning}.
For each episode $i$ starting from $s$ being the MA's initial
state~$s_0$ assumed w.l.o.g.\ to be probabilistic, the algorithm selects with
probability $\epsilon_i$ whether to \emph{explore} possibly new
decisions by sampling uniformly from the set $P(s)$ or to \emph{exploit} what has been learned so far in the
Q-table (line~\ref{alg:q-learning:exex}).
This is called an \emph{$\epsilon$-greedy strategy}.
Afterwards, the next state is selected randomly according to the
transition's probability distribution and the reward is
collected (lines \ref{alg:q-learning:xs} and
\ref{alg:q-learning:rew}).
Since the Q-table is only defined on states with probabilistic transitions, all rewards of states with only Markovian transitions that follow up to the next probabilistic one are accumulated into the Q-table entry of the preceding probabilistic state (lines \ref{alg:q-learning:while}-\ref{alg:q-learning:elihw}).
%
%
After collecting the rewards, the Q-table is updated
(line~\ref{alg:q-learning:learn}).  Here, the learning factor
$\alpha_i$ determines the impact of new information on the existing
knowledge.
Typically, $\alpha_i$ and $\epsilon_i$ decrease as $i$---the number of
episodes run so far---increases.
A higher $\epsilon_i$ allows the algorithm to explore various actions,
avoiding premature convergence to sub-optimal solutions. As the
algorithm learns more about the environment, it becomes beneficial to
gradually reduce $\epsilon_i$, leading to a focus on exploiting the
best-known strategies.  Similarly, a high initial $\alpha_i$ allows
for rapid learning but can destabilise due to noisy data or
outliers. Reducing $\alpha_i$ over time helps stabilise the learning
process as the algorithm converges, integrating new information more
conservatively~\cite{WD92}.  As an optimisation, we skip
Q-table updates for states with only one transition.

\paragraph{Discounting and convergence.}
An episode is very similar to a simulation run. 
The main differences to simulation as used for LSS are that we update
the Q-table to estimate the ``quality'' $\mathcal{Q}(s, a)$ of
taking the action $a$ from state $s$ and follow an
``$\epsilon_i$-greedy'' strategy. 
RL traditionally optimises for expected discounted rewards, thus the discount factor $\gamma$.
Since our objective is undiscounted, we set $\gamma$ to $1$.
Standard results~\cite{WD92,SB98} guarantee that the algorithm
converges to the maximal expected accumulated reward
as $\nrepisodes\to\infty$, as long as
\begin{enumerate*}[(i)]
\item%
  every state is guaranteed to be visited infinitely often (\ie
  $\epsilon_i>0$ for all $i\geq 0$),
\item%
  parameters $\alpha_i$ and $\epsilon_i$ decrease as
  $i\to\infty$, and
\item%
  the families $\{\alpha_i\}_{i\geq 1}$ and $\{\epsilon_i\}_{i\geq 1}$
  fulfill some variant of the stochastic approximation conditions.
\end{enumerate*}

\subsubsection{Performance and scalability.}

While QL is similar to LSS in that it uses simulation runs---so both are so-called model-free techniques---its memory
usage is in $\mathcal{O}(|\St| \cdot |\Act|)$ (for the
Q-table) and thus more similar to probabilistic model checking.
For many models, however, QL only explores---and thus stores a
Q-value for---a subset of $\St$.  This happens because some parts of
the state space have a very low probability of being reached from
$s_0$ within the specified number of episodes.  Additionally, no
Q-values need to be stored for states that have Markovian transitions only or just a single probabilistic transition.
The time spent in LSS and QL depends on the number of
simulation runs performed.  For LSS using our smart sampling approach,
we need $\mathcal{O}(N \cdot \log{K})$ runs (where $K$ is the strategy
budget and $N$ is the simulation budget per iteration), while
QL needs $\mathcal{O}(\nrepisodes)$ runs (where $\nrepisodes$
is the number of episodes to learn from).  Each run (episode) in
QL is however slightly more computationally expensive than in
LSS due to the computations involving the Q-table.

\section{Tool Extensions}
\label{sec:ToolExtensions}

To tackle our new open-pit mining case study, we extended the \modes statistical model checker~\cite{BDHS20} of the \toolset~\cite{HH14} with three crucial new features that we describe in this section:
Syntax and LSS- as well as QL-support for partial obervability (\Cref{sec:PartialObservability}), a constant-memory implementation of strategy extraction from SMC (\Cref{sec:StrategiesFromSMC}), and a connection to the \dtcontrol tool to obtain a decision tree representation of strategies (\Cref{sec:ToolExtensionDecisionTrees}).
We also implemented LSS and QL for MA and memoryless strategies as described in the previous section, as QL was previously only supported for MDP.

\subsection{Partial Observability}
\label{sec:PartialObservability}

In practical machine learning, it is common practice to expose to the learner not the full state of a model, with values for all of the model's variables, but only a set of carefully selected \emph{features}.
The learner then effectively works on a smaller state space.
We can cast feature selection as a variant of \emph{partial observability}, though not with the intention of finding the optimal strategy for the actual partially-observable MDPs~\cite{KLC98} or MAs:
such strategies require tracking probability distributions about which actual states the model could be in based on the history of observations, resulting in complex strategies and complicating any possible LSS or QL approach.

In the feature-oriented approach, we simply replace states by observables in the LSS and QL algorithms.
Let $\omega\colon \St \to \mathit{Obs}$ map states to some finite set of observables $\mathit{Obs}$.
Then, in \Cref{alg:lss-sim} for LSS, we replace line~\ref{alg:lss-sim:else1} by\\[2pt]
\centerline{$
\tuple{a, \mu} := (\mathcal{H}(\sigma.\omega(s)) \mathbin\mathrm{mod} |P(s)|)\text{-th element of }P(s);
$}\\[2pt]
in \Cref{alg:q-learning} for QL, we replace line~\ref{alg:q-learning:xs} by
$
\oplus_{\epsilon_i}
\displaystyle
\!\!\mathop{\textrm{arg max}}_{\tuple{a'\!,\mu'}\in P(s)}\!\! \mathcal{Q}(\omega(s),a')
$
and line~\ref{alg:q-learning:learn} by\\[1pt]
\centerline{$
\mathcal{Q}(\omega(s),a) := (1-\alpha_i) \cdot \mathcal{Q} (\omega(s),a)
           +
            \alpha_i \cdot \left(r + \gamma\cdot\max_{\tuple{a',\mu'}\in P(s')} \mathcal{Q}(\omega(s'),a')\right).
$}\\[3pt]
We require that $\omega$ projects only states with the same actions to the same observable, \ie
\begin{equation}
\label{eq:ConsistentPO}
\forall\, o \in \mathit{Obs}\ \forall\, s_1, s_2 \in \omega^{-1}(o)\colon (\exists\, \mu_1 \colon s_1 \xtrp{a} \mu_1) \Leftrightarrow (\exists\, \mu_2 \colon s_2 \xtrp{a} \mu_2).
\end{equation}
Otherwise, the Q-table would become inconsistent; for LSS, the inconsistencies would become apparent during strategy extraction (see \Cref{sec:StrategiesFromSMC} below).

To allow the modeller to describe $\mathit{Obs}$ and $\omega$, we extended the \modest language by the ability to mark a subset of the model's variables as \lstinline|observable|.
Then $\mathit{Obs}$ is the set of states projected to observable variables only, and $\omega$ discards the values of all non-observable values in a state.
This is inspired by the way the \lang{Prism} language supports partially-observable models~\cite{NPZ17}.

\subsection{Strategies from SMC}
\label{sec:StrategiesFromSMC}

\begin{figure}[t]
\centering
\begin{tikzpicture}[on grid,auto]
  \node[draw,rounded corners,align=center] (t1) {\strut simulation\\thread$_1$};
  \node[right=1.25 of t1] (tdots) {$\cdots$};
  \node[draw,rounded corners,align=center,right=1.25 of tdots] (tn) {\strut simulation\\thread$_n$};
  \node[draw,align=center,below=1.5 of t1] (f1) {\strut strategy\\file$_1$};
  \node[below=1.5 of tdots] (fdots) {$\cdots$};
  \node[draw,align=center,below=1.5 of tn] (fn) {\strut strategy\\file$_n$};
  \node[below=0.75 of tdots,align=center] (rr) {\textit{state-action}\\[-1.5pt]\textit{pairs}};
  \node[draw,rounded corners,align=center,right=2.5 of tn] (se) {\strut statistical\\\strut evaluation};
  \node[right=2.5 of se] (res) {\textit{\bfseries estimate}};
  \node[draw,align=center,below=1.5 of f1] (mf) {\strut merged\\\strut strategy file};
  \node[draw,align=center,right=5 of mf] (tf) {\strut strategy\\\strut table file};
  \node[draw,rounded corners,right=2.5 of tf] (dtc) {\strut\dtcontrol};
  \node[above=2.25 of dtc,align=center] (dt) {\textit{\bfseries\strut decision}\\[-1.5pt]\textit{\bfseries\strut tree}};
  \node[above=2.68 of dtc] (plus) {\textit{\bfseries +}};

  \node[above right=0.95 and 1.25 of tn] (rr) {\textit{run results}};
  \coordinate[above=0.75 of t1] (t1x);
  \coordinate[above=0.75 of tn] (tnx);
  \coordinate[above=0.75 of se] (sex);
  \node[below=0.65 of f1,anchor=west] (cc) {concatenate};
  \coordinate[below=0.825 of f1] (f1x);
  \coordinate[below=0.825 of fn] (fnx);

  \path[-latex]
    (t1) edge[transform canvas={xshift=-3mm}] (f1)
    (t1) edge[transform canvas={xshift=-1mm}] (f1)
    (t1) edge[transform canvas={xshift=1mm}] (f1)
    (t1) edge[transform canvas={xshift=3mm}] (f1)
    (tn) edge[transform canvas={xshift=-3mm}] (fn)
    (tn) edge[transform canvas={xshift=-1mm}] (fn)
    (tn) edge[transform canvas={xshift=1mm}] (fn)
    (tn) edge[transform canvas={xshift=3mm}] (fn)
    (se) edge (res)
    (mf) edge node[above,inner sep=0.35pt] {\strut external-memory} node[below,inner sep=1.5pt] {\strut sort and deduplicate} (tf)
    (tf) edge (dtc)
    (dtc) edge (dt)
  ;
  \draw[-latex] (t1) -- (t1x) -- (sex) -- (se);
  \draw[] (tn) -- (tnx);
  \draw[-latex] (f1) -- (mf);
  \draw[] (fn) -- (fnx) -- (f1x);

  \coordinate[above left=1.2 and 1 of t1] (d1);
  \coordinate[below left=0.5 and 1 of f1] (d2);
  \coordinate[below right=2.0 and 1 of se] (d3);
  \coordinate[above right=1.2 and 1 of se] (d4);
  \draw[dotted] (d1) -- (d2) -- (d3) -- (d4) -- (d1);
  \coordinate[below left=0.8 and 1 of mf] (d5);
  \coordinate[below right=0.8 and 1 of dtc] (d6);
  \coordinate[below right=2.0 and 1 of res] (d7);
  \draw[dotted] (d2) -- (d5) -- (d6) -- (d7) -- (d3);
  \node[above=0 of d3,anchor=south east] (smc) {SMC};
  \node[below=0 of d6,anchor=south east] (pp) {postprocessing};
  \coordinate[left=1 of t1] (d12);
  \coordinate[left=3.0 of t1] (d12x);
  \path[-latex]
    (d12x) edge node[above,inner sep=0.35pt,pos=0.45] {\strut\textit{\bfseries MA model}} node[below,inner sep=2.0pt,pos=0.425,align=center] {\strut \textit{\bfseries\strut+\,strategy id}\\[-0.5pt]\textit{\strut\bfseries\phantom{+}or Q-table}} (d12)
  ;
\end{tikzpicture}
\caption{Schematic overview of strategy extraction from SMC}
\label{fig:StrategyExtraction}
\end{figure}
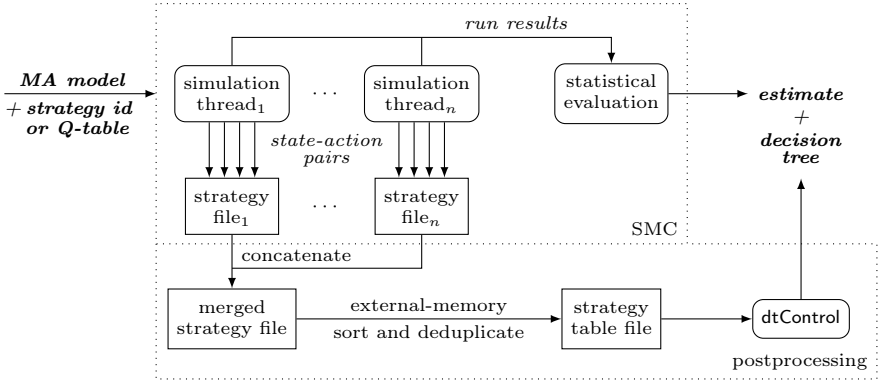

When the LSS process with smart sampling of \Cref{alg:lss} completes, it returns a strategy in the form of a strategy identifier $\sigma_{\max} \in \ZZ_{32}$.
When QL as in \Cref{alg:q-learning} is finished, it returns a strategy in the form of the Q-table $\mathcal{Q}$.
Both representations are not very useful:
$\sigma_{\max}$ contains no explicit information about its expected reward or the actual decisions the strategy makes to attain it, while $\mathcal{Q}$ is large, contains the decisions only implicitly, and $\max_{\tuple{a,\mu}\in P(s_0)} \mathcal{Q}(s_0,a)$ is an approximation of the expected reward with no guarantee as to how far from the strategy's real expected reward it is.
To obtain an estimate of the strategy's expected reward with a statistical error guarantee, \modes therefore performs a separate SMC analysis of the MA under the strategy.
That is, it performs a number of new simulation runs that is sufficient to attain the error and confidence requested by the user, and returns their average value as the estimate of the expected reward.

For the open-pit mining case study, however, we are also interested in understanding the corresponding strategy:
How does the mine operator need to schedule its trucks in order to achieve the computed (hopefully near-maximal) total load of material transported?
To answer this question, we have extended \modes with a novel method to extract (an approximation of) the strategy in a more explicit form that runs in constant memory, just like SMC and LSS.
\Cref{fig:StrategyExtraction} shows a schematic overview of the implementation:
Each simulation thread (in a multi-core or distributed setting) writes all state-action pairs chosen by the strategy that it encounters during the simulation runs it performs into a separate binary file on disk.
Thus there is no overhead for coordinating the threads while they run.
Only after the SMC process is done and the estimate has been returned do we process these files:
We first concatenate them into a single binary ``merged'' strategy file.
Then we apply an external-memory merge sort algorithm to sort the state-action pairs in this file according to some arbitrary order for the purpose of eliminating any duplicates.
At this point, we detect if an LSS strategy under partial observability violates \Cref{eq:ConsistentPO}.  If this is the case, the error is reported and the process is aborted.
Otherwise, the resulting table of action choices for unique states is written to a text file in \tool{mcsta}'s strategy file format.


\subsection{Decision Trees}
\label{sec:ToolExtensionDecisionTrees}

The tabular strategy file that our new SMC strategy extraction method in \modes delivers is human-readable and arguably more useful that the integer strategy identifier or the huge Q-table.
It may still be very large, however, hiding interesting patterns such as dependencies between ranges of state variable values and strategy choices.
To obtain a more explainable representation, we implemented a new connection from \modes and \tool{mcsta} to the \dtcontrol tool.
\dtcontrol~\cite{AJKWWY21} reads tabular strategy representations produced by model checkers such as \tool{Prism}~\cite{KNP11} and \tool{Storm}\cite{HJKQV22} or by \tool{Uppaal Stratego}~\cite{DJLMT15} and learns a decision tree that succinctly represents the strategy.
We implemented support for the textual strategy file format now used by \modes and \tool{mcsta} via a new dataset loader for \dtcontrol.
By supporting both the SMC and the PMC tool of the \toolset, we can also compare---on small models---the trees \dtcontrol learns for the complete strategy obtained by \mcsta vs.\ the one for the usually incomplete table (that only contains the states actually visited during SMC) generated by \modes.
We show decision trees obtained via this new connection for the open-pit mining case study in \Cref{sec:ExperimentalResults}.

\section{Case Study Modelling}
\label{sec:ModellingAndAnalysis}

To improve the tractability of the truck dispatching problem, we make some simplifications on the open-pit mine model.
First, we assume all timing aspects stochastically distributed under exponential distributions, enabling modelling as MA.
Though different truck models can haul significantly different loads in practice, we assume they are all equal.
Also, distances, which determine the travelling times, are point to point.
However, to reduce the number of combinations, we consider that the distance towards one point---a shovel or a dump---is the same from any other point (but may differ from the distance towards a different~point).

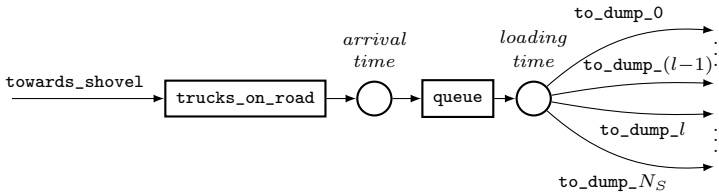
\begin{figure}[t]
  \centering
  \begin{tikzpicture}[on grid,auto,align at top]
    \node (n0) {};
    \node[waiting] (n1) [right=3.2 of n0] {\rule[-.4ex]{0pt}{.7em}\texttt{trucks\_on\_road}};
    \node[server]  (n2) [right=1.7 of n1] {};
    \node[waiting] (n3) [right=1.1 of n2] {\rule[-.4ex]{0pt}{.7em}\texttt{queue}};
    \node[server]  (n4) [right=1 of n3] {};
    \node (n5)  [above right=0.9  and 2.5 of n4] {};
    \node (n6)  [above right=0.7  and 2.4 of n4] {$\vdots$};
    \node (n7)  [above right=0.2  and 2.5 of n4] {};
    \node (n8)  [below right=0.2  and 2.5 of n4] {};
    \node (n9)  [below right=0.45 and 2.4 of n4] {$\vdots$};
    \node (n10) [below right=0.9  and 2.5 of n4] {};
    \node (n11) [above=0.8 of n2] {\textit{arrival}};
    \node (n12) [above=0.5 of n2] {\textit{time}};
    \node (n13) [above=0.8 of n4] {\textit{loading}};
    \node (n14) [above=0.5 of n4] {\textit{time}};

    \path[-latex]
      (n0) edge node[xshift=-5] {\texttt{towards\_shovel}} (n1)
      (n1) edge node {} (n2)
      (n2) edge node {} (n3)
      (n3) edge node {} (n4)
      (n4) edge[bend left=23]  node[above,xshift=-2,yshift=3] {\texttt{to\_dump\_0}} (n5)
      (n4) edge[bend left=6]  node[above,xshift=6] {\texttt{to\_dump\_{$(l{-}1)$}}} (n7)
      (n4) edge[bend right=6] node[below,xshift=4] {\texttt{to\_dump\_{$l$}}} (n8)
      (n4) edge[bend right=23] node[below,xshift=-4,yshift=-3] {\texttt{to\_dump\_{$N_S$}}} (n10)
    ;
  \end{tikzpicture}
  \caption{Schematic view of the behaviour of trucks in a \texttt{ShovelSystem}}\label{fig:shovel-system}
\end{figure}

\begin{figure}[t]
\begin{lstlisting}[numbers=left]
process ShovelSystem(int(0..§$N_S$§) shovel_id, int(0..MAX_TIME) t_time) {

   int(0..NR_TRUCKS) queue = 0;§\label{ln:startvar}§
   int(0..NR_TRUCKS) trucks_on_road = 0; // trucks travelling to this shovel
   observable int(0..MAX_OBS) stress = 0; // shovel stress level
   observable bool full = false; // if true, a truck was just loaded§\label{ln:endvar}§


   do {§\label{ln:startloop}§
   :: towards_shovel // a truck is sent towards this shovel§\label{ln:truckstarts}§
      {= trucks_on_road++, stress = min(trucks_on_road+queue, MAX_OBS) =}

   :: when(trucks_on_road > 0) // a truck arrives to this shovel§\label{ln:arrivalcond}§
      rate(trucks_on_road / t_time) {= queue++, trucks_on_road-- =}§\label{ln:arrival}§

   :: when (queue > 0) rate(1 / LOAD_TIME) // a truck is being loaded§\label{ln:loadcond}§
      {= queue--, stress = min(trucks_on_road + queue, MAX_OBS), full = true =}§\label{ln:load}§

      // The truck is sent to a dump of the right kind for this shovel
   :: when(full && shovel_id <  k) to_dump_0        §\hspace{3.3pt}§{= full = false =}§\label{ln:startdispatch}§
      §$\ldots$§
   :: when(full && shovel_id <  k) to_dump_§\color{red!50!black}$(l-1)$§   §\hspace{2.9pt}§{= full = false =}
   :: when(full && shovel_id >= k) to_dump_§\color{red!50!black}$l$§         §\hspace{0.75pt}§{= full = false =}
      §$\ldots$§
   :: when(full && shovel_id >= k) to_dump_§\color{red!50!black}$(N_D-1)$§ §\hspace{0.3pt}§{= full = false =}§\label{ln:enddispatch}§
   }§\label{ln:endloop}§
}

process DumpSystem(int(0..§$N_D$§) dump_id, int(0..MAX_TIME) h_time) {

   int(0..NR_TRUCKS) queue = 0;
   int(0..NR_TRUCKS) trucks_on_road = 0; // trucks travelling to this dump
   observable int(0..MAX_OBS) stress = 0; // dump stress level
   observable bool empty = false; // if true, a truck has just unloaded


   do {
   :: towards_dump // a truck is sent towards this dump
      {= trucks_on_road++, stress = min(trucks_on_road+queue,MAX_OBS) =}

   :: when (trucks_on_road > 0) // a truck arrives to this dump
      rate(trucks_on_road / h_time) {= queue++, trucks_on_road-- =}

   :: when (queue > 0) rate(1 / UNLOAD_TIME) // a truck is dumping the material
      {= queue--, stress = min(trucks_on_road + queue, MAX_OBS),
         load = TRK_LOAD, empty = true =}§\label{ln:reward}§

      // The truck is sent to a shovel
   :: when (empty) to_shovel_0        §\hspace{1.75pt}§{= empty = false =}
      §$\ldots$§
   :: when (empty) to_shovel_§\color{red!50!black}$(N_S-1)$§ {= empty = false =}
   }
}
\end{lstlisting}
\caption{\modest models for the shovel and dump systems}
\label{fig:model}
\end{figure}

\paragraph{Modest model.}
The schematic of all activities that involve
a truck---from starting to go to the shovel until it is
finally served and assigned to a dump---is depicted in
\Cref{fig:shovel-system}.  This is what the \modest model of the
shovel system in \Cref{fig:model} aims to capture.
The process \lstinline{ShovelSystem} receives two parameters:
\lstinline{shovel_id} identifies this shovel and determines if it loads
useful or waste material while \lstinline{t_time} is the average time
that a truck needs to travel to this shovel.
Variables are introduced in lines \ref{ln:startvar}-\ref{ln:endvar}.
Variable \lstinline{trucks_on_road} counts the trucks that are
approaching this shovel and variable \lstinline{queue} counts the
trucks that have arrived and are queuing for the shovel.  The
observable Boolean \lstinline{full} indicates when a truck is
full and should be dispatched to some dump.  Observable variable
\lstinline{stress} do not play a role in the control flow; we explain it later.

The whole behaviour is captured in the loop in lines
\ref{ln:startloop}-\ref{ln:endloop}.  When a truck is dispatched to
the shovel, it synchronises with action \lstinline{towards_shovel}
(line~\ref{ln:truckstarts}) which in turn increases the counter of
trucks travelling towards this shovel.
If there are trucks travelling to the shovel
(line~\ref{ln:arrivalcond}), then one of them may arrive.
The time of arrival of a truck is determined by an
exponential distribution of average \lstinline{t_time} that is weighted
by the total number of trucks approaching the shovel
(line~\ref{ln:arrival}). On arrival, the truck enqueues
and the number of approaching trucks decreases by one.
If some truck is waiting for loading
then there is certainly one being loaded and the time of loading is
determined by an exponential distribution of average
\lstinline{LOAD_TIME} (line~\ref{ln:load}).  When the truck finishes
loading, it is removed from the queue and variable \lstinline{full} is
set to indicate that the truck needs to be dispatched.  This happens
in lines \ref{ln:startdispatch}-\ref{ln:enddispatch}. In particular,
if \lstinline{shovel_id} is smaller than a given number $k$, this shovel
loads useful material and the truck can only be dispatched to one of
the $l$ dumps with stockpiles.  If instead \lstinline{shovel_id} is
greater or equal to $k$, it loads waste and hence the truck should be
dispatched to one of the $N_D-l$ waste dumps.
The dispatching is carried out with one of the \lstinline{to_dump_}{\color{red!50!black}$i$}
actions which in turns synchronises with the \lstinline|towards_dump|
action of the dump system with identification number $i$ (in process
\lstinline|DumpSystem|).  Action
\lstinline|towards_dump| in the dump is the analogon to
\lstinline|towards_shovel| in the shovel system.  A dump system
is modelled similarly, the only difference being
that trucks can be dispatched to any shovel regardless of the material
they handle.

To obtain a reduced observable domain, only variables \lstinline{full}
and \lstinline|stress| are made observable. Variable \lstinline{full} needs
to be observable in order to distinguish the states in which the
dispatching actions are enabled to ensure \Cref{eq:ConsistentPO} holds.  Variable
\lstinline|stress| is specifically designed to have a reduced observable
domain and can be understood as the shovel stress level.  It contains
the total number of trucks that are either approaching to or waiting
in the shovel but up to the maximum \lstinline{MAX_OBS} which indicates
the maximum stress.  This constant is different in each shovel; we
made it depend on the arrival time and the loading time.  A similar
decision has been taken for the dump systems.

Apart from the shovel systems and the dump systems, the model is
completed with an initialisation process.  Since the mine is not
assumed to be in any particular initial state, the initialisation
module dispatches the trucks nondeterministically to any of the sites.


\paragraph{Optimisation objective.}
The objective is to maximise productivity, \ie the expected total load of material moved from the shovels
to the dumps during a shift.  Hence, we obtain a branch reward of the capacity
of the truck each time a truck finishes dumping its load.  This is
indicated with the assignment
\lstinline{load = TRK_LOAD} in line \ref{ln:reward} of the
\lstinline{DumpSystem} (see \Cref{fig:model}) to the \emph{transient} global variable \lstinline{load}.
A transient variable is not part of
the state and only takes a value other than its default ($0$ for \lstinline{load}) during the execution of the assignment block.
%
In \modest notation, we analyse the property\\[1pt]
\centerline{\strut\lstinline{Xmax[T == SHIFT](S(load))}}\\[1pt]
which represents the maximum
expected value of the sum of \lstinline{load} values along the execution
before reaching time \lstinline{SHIFT}, which we set to $200$.

Though not particularly interesting for the truck dispatching problem,
we also estimate the \emph{minimum} expected accumulated value of
\lstinline{load} 
as well as the value obtained by the uniform random strategy (that randomly chooses a transition every time it encounters a state with multiple probabilistic transitions).
This is useful to gain insights into the ability of the LSS and QL engines to actually optimise (measured by the spread between maximum and minimum value) and find nontrivial strategies (that differ from the uniform random one).
In addition, as the result of the integration of \modes
with \dtcontrol, we are able to derive decision trees that explain
the near-optimal strategies that we find. 

\section{Experimental Results}
\label{sec:ExperimentalResults}

\begin{table}[t]
  \vspace{-4ex}
  \centering\smaller[2]
  
  \caption{Experimental setup}
  \label{tab:experiments}
  \vspace{2ex}
  
  \begin{subtable}[b]{.56\linewidth}
    \def\NUM#1{\texttt{\#}\,#1}
    \centering
    \begin{tabular}{r@{:\quad}r@{~~}r@{~~}r@{~~}r@{~~}r@{~~}r@{~~}r}
    \toprule
    \bfseries \NUM{trucks}
    & \bfseries 4 & \bfseries 5 & \bfseries 9 & \bfseries 10 & \bfseries 35 & \bfseries 40 & \bfseries 80
    \\\midrule
    \NUM{shovels}
    & 6 & 1 & 3 &  6 &  6 &  8 & 10
    \\[.2ex]
    \NUM{dumps}
    & 5 & 2 & 2 &  5 &  5 &  8 & 10
    \\[.2ex]
    \NUM{ore shovels}
    & 3 & 0 & 1 &  3 &  3 &  4 &  5
    \\[.2ex]
    \NUM{ore dumps}
    & 3 & 0 & 1 &  3 &  3 &  4 &  5
    \\[.5ex]
    \arrayrulecolor{gray}\midrule[.5pt]
    combinations\, shovel $\rightarrow$ dump
    & 15 & 2 & 3 & 15 & 15 & 32 & 50
    \\
    combinations\, shovel $\leftarrow$ dump
    & 30 & 2 & 6 & 30 & 30 & 64 & 100
    \\
    comb.~ore shovel $\rightarrow$ ore dump
    & 9 & 0 & 1 & 9 & 9 & 16 & 25 
    \\
    \arrayrulecolor{black}\bottomrule
    \end{tabular}
    \caption{Mine model instances---name is \NUM{trucks}}
    \label{tab:models}
  \end{subtable}
  \hfill
  \begin{subtable}[b]{.38\linewidth}
    \centering
    \begin{tabular}{L{.35\linewidth}L{.63\linewidth}}
    \toprule
    \bfseries Run name & \bfseries Parameters
    \\\midrule
    LSS\,10k\,1k & 10k simulation traces, 1k strategies
    \\
    LSS\,100k\,10k & 100k simulation traces, 10k strategies
    \\
    Qlearn\,100k & 100k episo., 0.5 $\lambda$, 0.02 fin-$\lambda$, 1.0 $\varepsilon$, 0.02 fin-$\varepsilon$
    \\
    Qlearn\,3M & 3M episo., 0.6 $\lambda$, 0.01 fin-$\lambda$, 0.6 $\varepsilon$, 0.02 fin-$\varepsilon$
    \\\bottomrule
    \end{tabular}
    \caption{\modes executions}
    \label{tab:runs}
  \end{subtable}

  \vspace{-4ex}
\end{table}

\paragraph{Setup.}
We use the LSS and QL implementations of \modes to estimate the maximum and minimum expected accumulated load on seven instances of our model from \Cref{sec:ModellingAndAnalysis}---see \Cref{tab:models} for an overview of their configurations.
%
For each instance, besides the uniform run, we execute \modes with the algorithms from \Cref{tab:runs} and let it build confidence intervals of 1\% relative half-width in the final SMC analysis. 
For each model, property, and run configuration, we execute \modes in two modes: the default with fully-observable states (\FO), which considers all model variables observable, and with partially-observable states (\PO) following the \lstinline|observable| annotations in the model.
Each experiment is run until a confidence interval of the requested width is built for the property under study.
We use the center of the intervals to compare the minimum and maximum loads found by each run on each model, keeping track of the wall-clock runtime.
We execute our experiments on an AMD Ryzen 9 7950X3D system (16 cores) with 128 GB of RAM running 64-bit Ubuntu Linux 22.04.

\begin{figure}[t]
    \centering
    \def\WIDTH{.47\linewidth}

    \begin{subfigure}{\WIDTH}
        \centering
        \includegraphics[width=.9\linewidth]{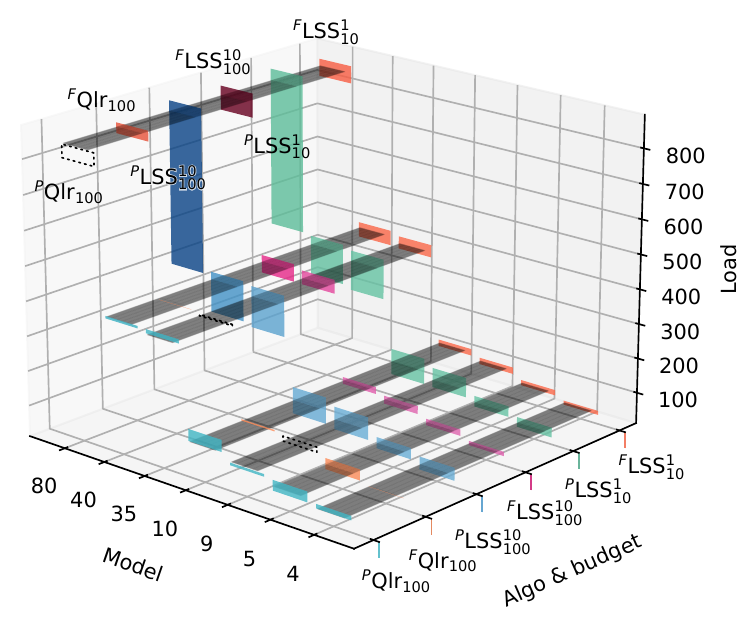}
    \end{subfigure}
    ~~
    \begin{subfigure}{\WIDTH}
        \centering
        \includegraphics[width=.9\linewidth]{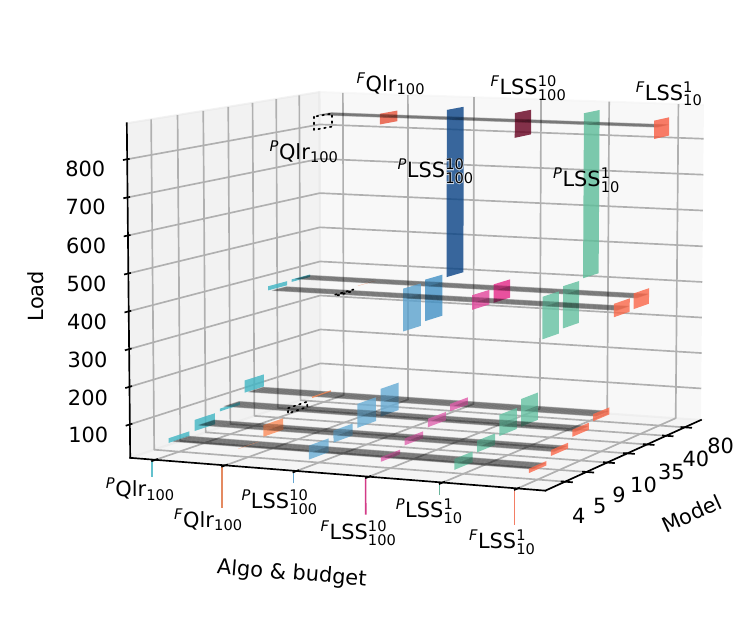}
    \end{subfigure}

    \caption{Overview of experimental results: minimum and maximum loads.}
    \label{fig:3D_plots}
\end{figure}

\begin{figure}
    \centering
    \def\WIDTH{.47\linewidth}

    \begin{minipage}{.49\linewidth}
        \centering Fully-observable states (\FO)
    \end{minipage}
    \begin{minipage}{.49\linewidth}
        \centering Partially-observable states (\PO)
    \end{minipage}
    \vspace{2ex}

    \textsmaller[2]{LSS: 10k simulation 1k strategies}\\[.5ex]
    \begin{subfigure}{\WIDTH}
        \centering
        \smaller
        \includegraphics[width=\linewidth]{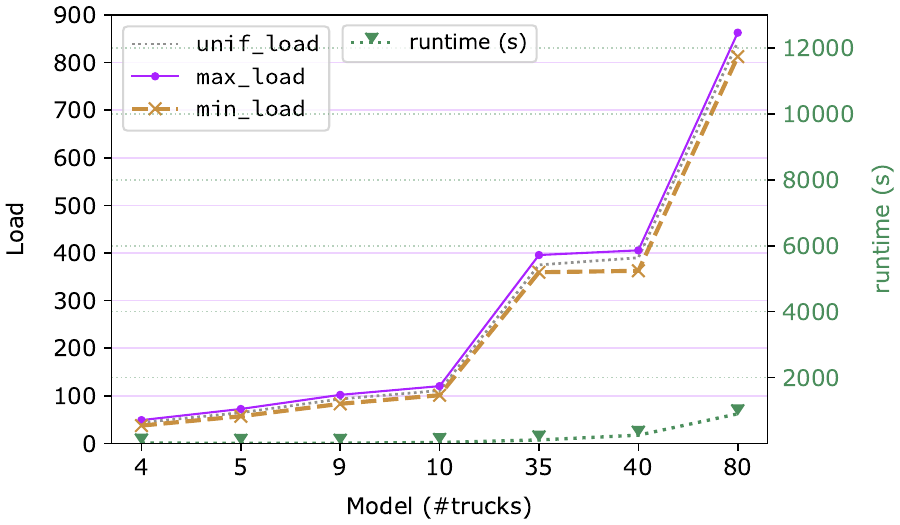}
    \end{subfigure}
    ~~
    \begin{subfigure}{\WIDTH}
        \centering
        \includegraphics[width=\linewidth]{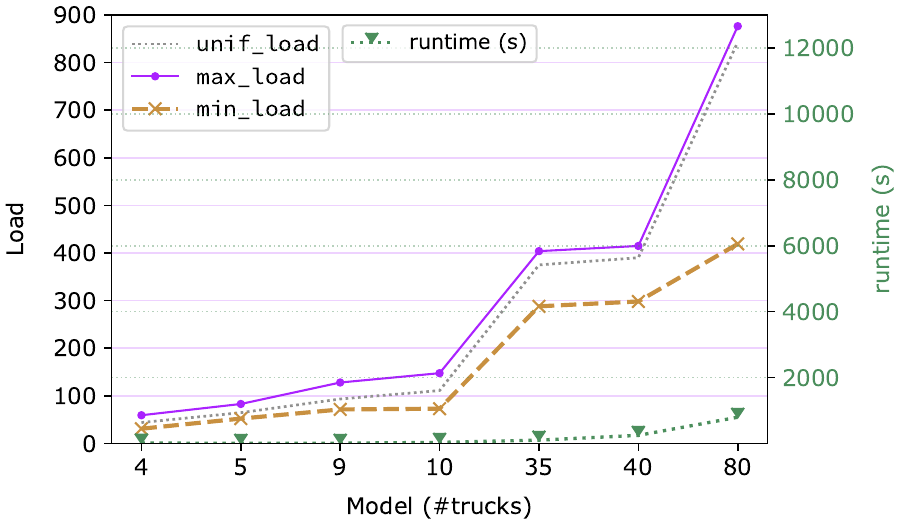}
    \end{subfigure}
    \vspace{1ex}

    \textsmaller[2]{LSS: 100k simulation 10k strategies}\\[.5ex]
    \begin{subfigure}{\WIDTH}
        \centering
        \smaller
        \includegraphics[width=\linewidth]{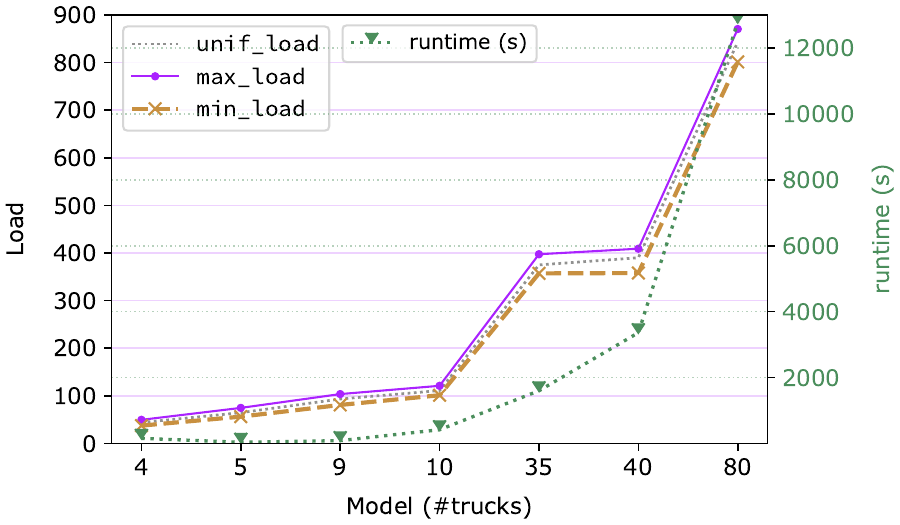}
    \end{subfigure}
    ~~
    \begin{subfigure}{\WIDTH}
        \centering
        \includegraphics[width=\linewidth]{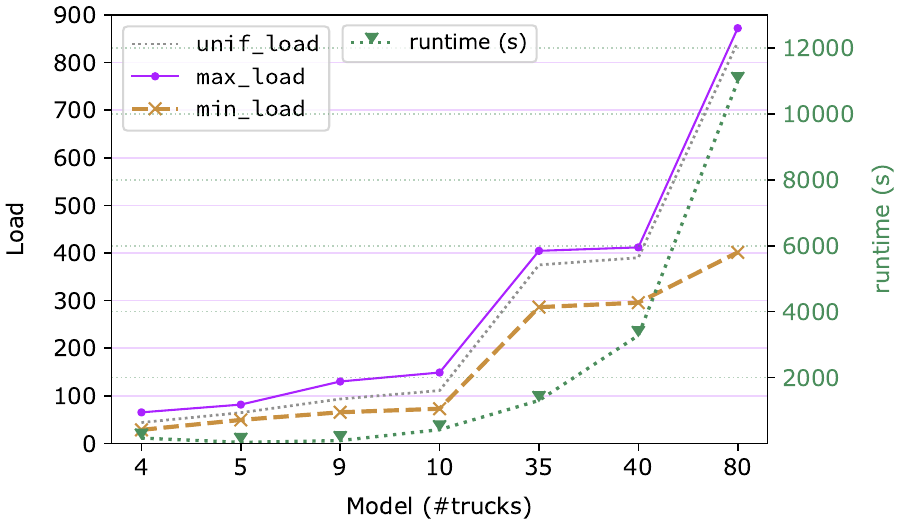}
    \end{subfigure}
    \vspace{1ex}

    \textsmaller[2]{Q-learning: 100k episodes}\\[.5ex]
    \begin{subfigure}{\WIDTH}
        \centering
        \smaller
        \includegraphics[width=\linewidth]{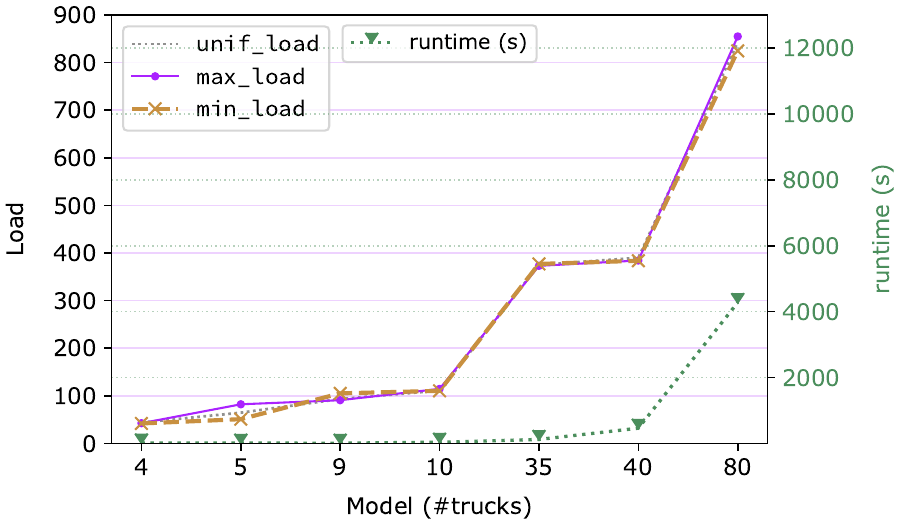}
    \end{subfigure}
    ~~
    \begin{subfigure}{\WIDTH}
        \centering
        \includegraphics[width=\linewidth]{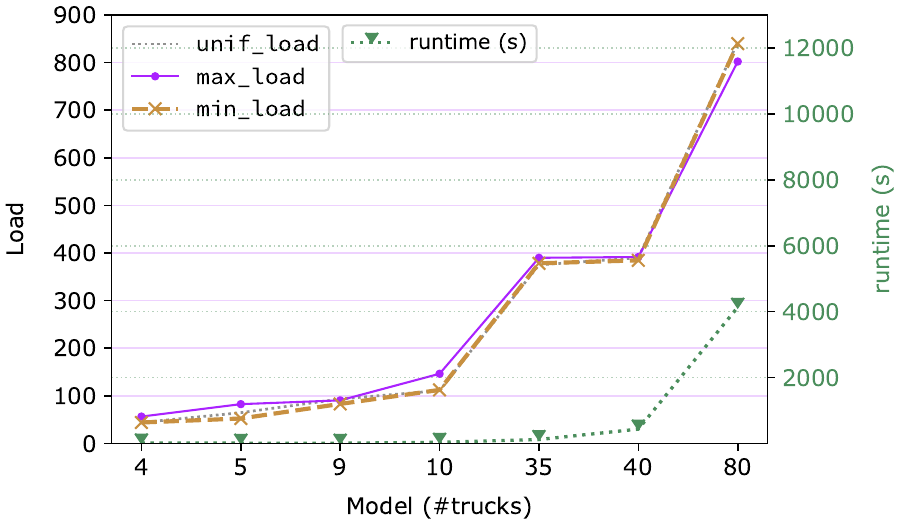}
    \end{subfigure}
    \vspace{1ex}

    \textsmaller[2]{Q-learning: 3M episodes}\\[.5ex]
    \begin{subfigure}{\WIDTH}
        \centering
        \smaller
        \includegraphics[width=\linewidth]{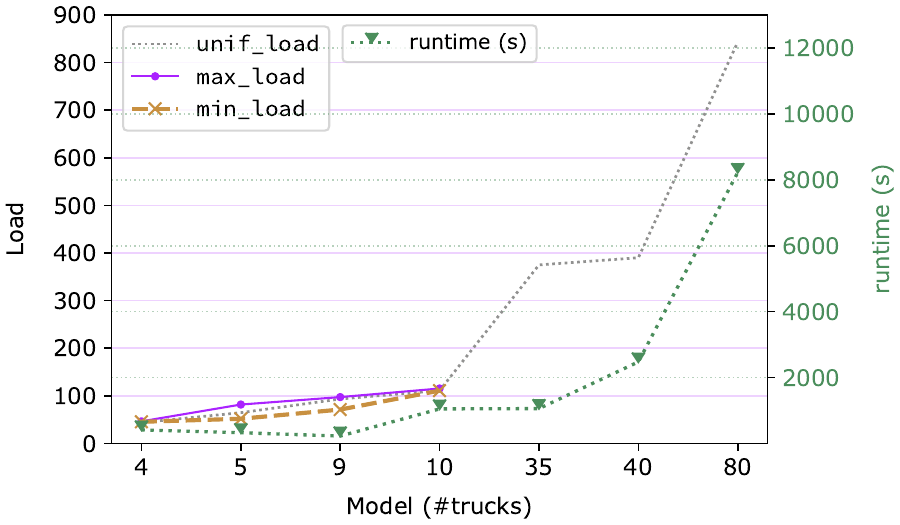}
    \end{subfigure}
    ~~
    \begin{subfigure}{\WIDTH}
        \centering
        \includegraphics[width=\linewidth]{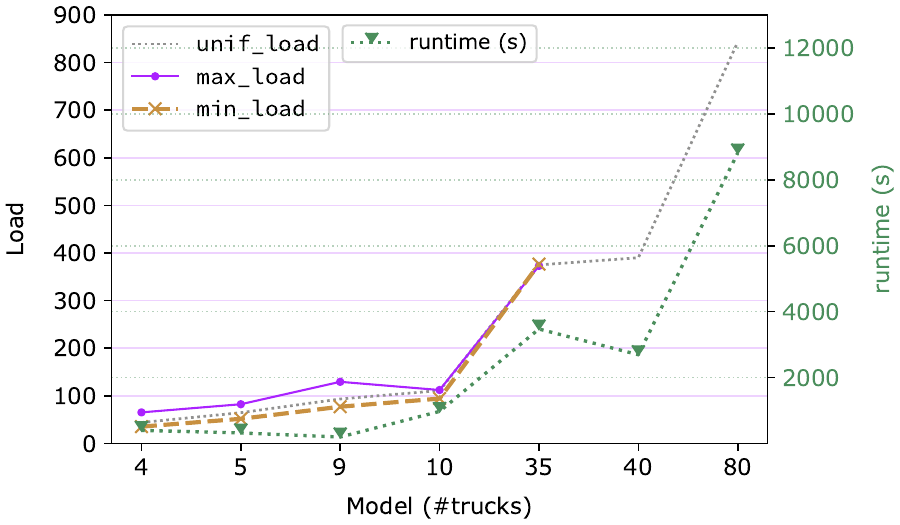}
    \end{subfigure}
    \vspace{1ex}


    \caption{Min.\ and max.\ expected loads via different algorithms and budgets}
    \label{fig:2D_plots}
\end{figure}

\paragraph{Results for loads.}
\Cref{fig:3D_plots,fig:2D_plots} show the results obtained in overview and in detail, respectively.
In \Cref{fig:3D_plots}, for each model and execution configuration, we plot a vertical rectangle: its lower bound is the minimum estimated by the configuration, and its upper bound is the maximum.
Therefore, the taller the rectangle, the bigger the difference between minimum and maximum.
The width of a rectangle bears no meaning, but its colour does: \PO runs are blue-greens and \FO runs are red-oranges; the darker the hue, the longer the runtime%
\footnote{%
We use the average runtime of both properties per model and run, whose coefficient of variation was in almost all cases below 10\% and in two cases 25.3\% and 25.1\%.
}---but see \Cref{fig:2D_plots} for an objective runtime comparison.

\Cref{fig:3D_plots} does not include results for Q-learning with the heavier budget (3M episodes), because these runs failed for all large models as shown in more detail in \Cref{fig:2D_plots}.
Furthermore, in three cases of Q-learning with 100k episodes the minimum load achieved was higher than the maximum load: this is indicated in \Cref{fig:3D_plots} with empty rectangles whose contour is drawn with a dashed line (\PO model 80, and \FO models 9 and 35).
This inversion never occurred when using LSS.
Finally, the transversal horizontal gray bars are the results of the uniform runs, 
which are plotted to serve as reference point.

\Cref{fig:3D_plots} shows how the uniform values are closer to the maximum load found by other runs than to the minimum loads.
Besides, all runs with the uniform random strategy took less than 2--3 seconds to finish.
However, in all cases for LSS, there was no overlap between the final confidence interval produced for the maximum load and that produced by the uniform random strategy.
Thus LSS shows a statistically significant improvement for minimum and maximum loads over the uniform strategy here, even in the lighter LSS configuration.

\Cref{fig:2D_plots} allows for a better comparison, where we see that the runtimes do not vary significantly between \FO and \PO runs, but instead what we gain by using partially-observable states are better minimising strategies.
\Cref{fig:2D_plots} also makes it clear that increasing the simulation budget of LSS from 10k\,1k to 100k\,10k impacts runtime severely, but has only a minor effect in terms of finding better strategies here.
Q-learning runs achieved no significantly different results than the uniform strategy, but incurred higher runtimes (except vs.\ LSS\,100k\,10k) and failed for all large models due to running out of memory for 3M episodes. 

\begin{figure}
    \centering
    \def\WIDTH{.47\linewidth}

    \begin{minipage}{.49\linewidth}
        \centering Fully-observable states (\FO)
    \end{minipage}
    \begin{minipage}{.49\linewidth}
        \centering Partially-observable states (\PO)
    \end{minipage}
    \vspace{2ex}

    \textsmaller[2]{LSS: 10k simulation 1k strategies}\\[.5ex]
    \begin{subfigure}{\WIDTH}
        \centering
        \smaller
        \includegraphics[width=\linewidth]{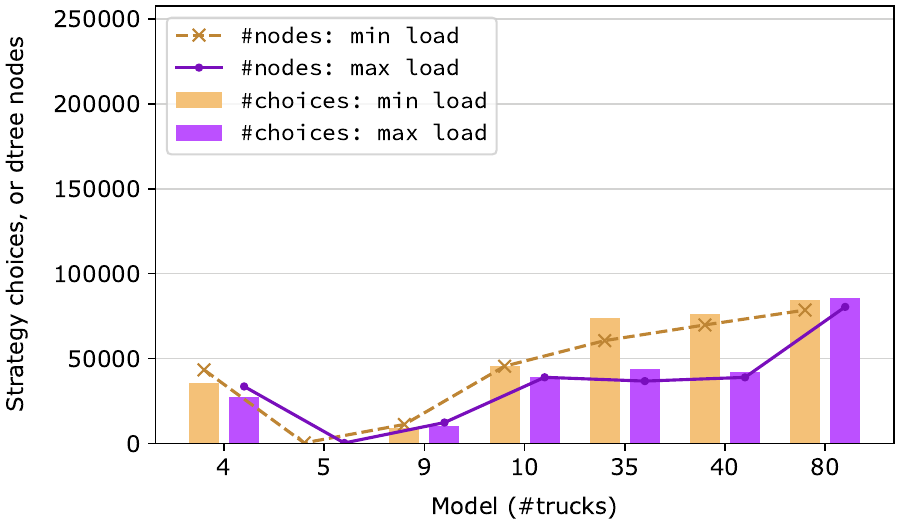}
    \end{subfigure}
    ~~
    \begin{subfigure}{\WIDTH}
        \centering
        \includegraphics[width=\linewidth]{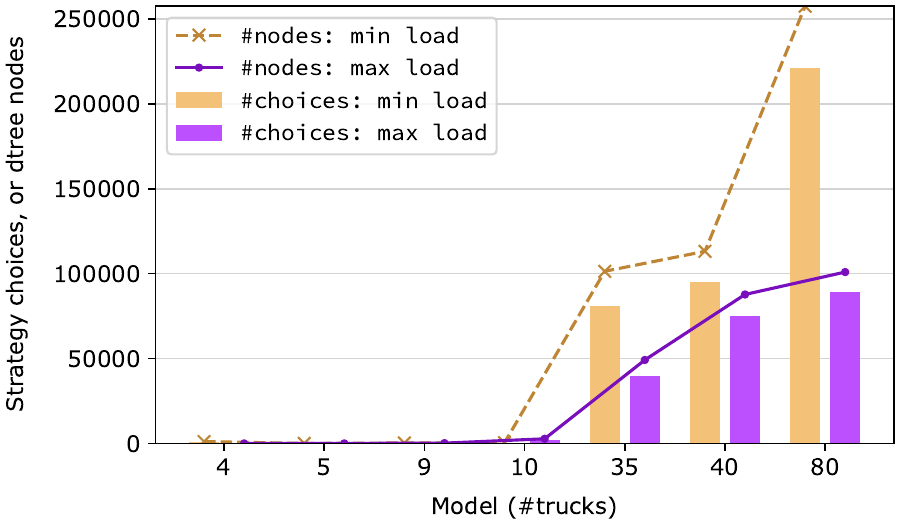}
    \end{subfigure}
    \vspace{1ex}

    \textsmaller[2]{LSS: 100k simulation 10k strategies}\\[.5ex]
    \begin{subfigure}{\WIDTH}
        \centering
        \smaller
        \includegraphics[width=\linewidth]{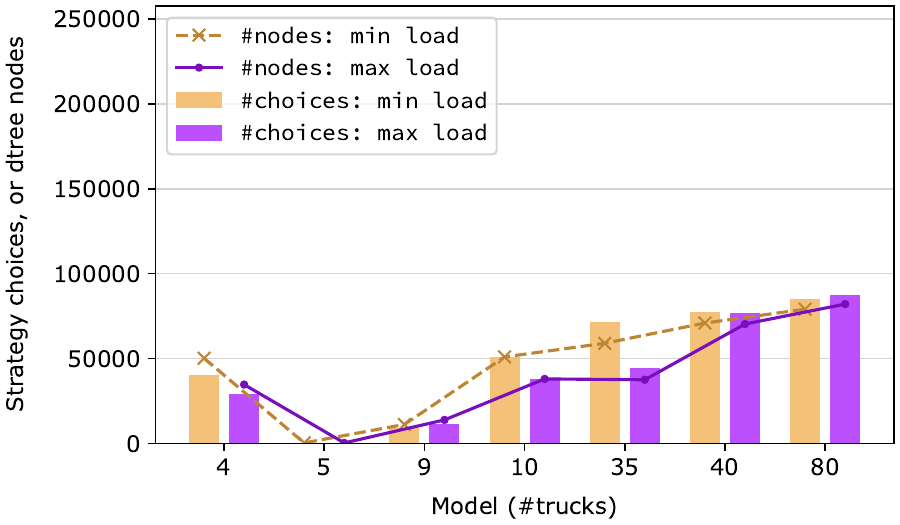}
    \end{subfigure}
    ~~
    \begin{subfigure}{\WIDTH}
        \centering
        \includegraphics[width=\linewidth]{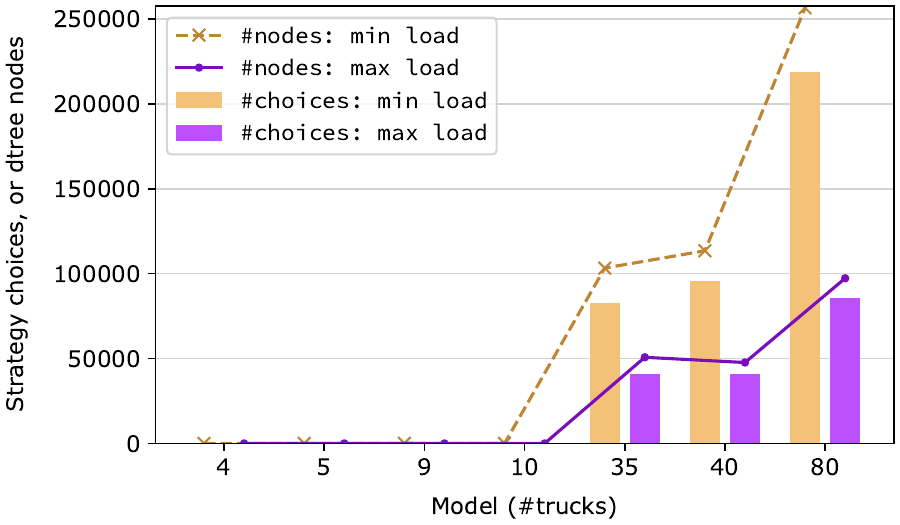}
    \end{subfigure}
    \vspace{1ex}

    \textsmaller[2]{Q-learning: 100k episodes}\\[.5ex]
    \begin{subfigure}{\WIDTH}
        \centering
        \smaller
        \includegraphics[width=\linewidth]{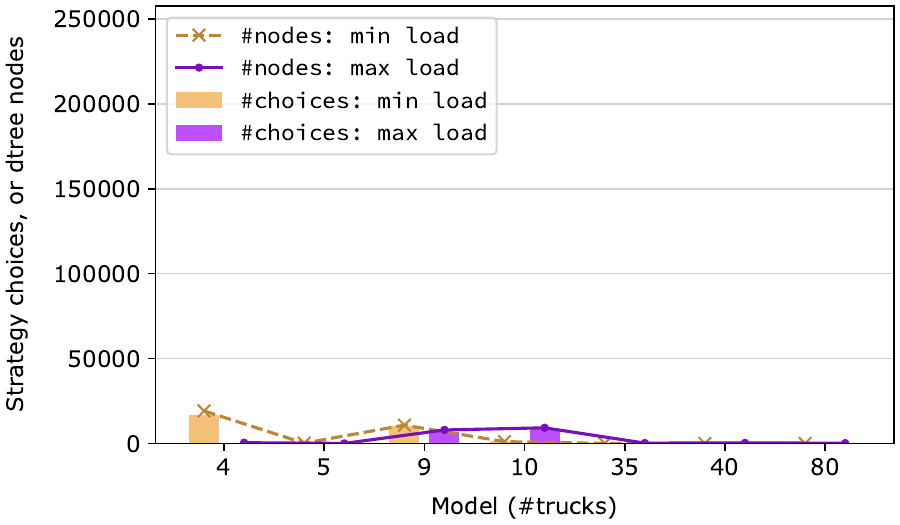}
    \end{subfigure}
    ~~
    \begin{subfigure}{\WIDTH}
        \centering
        \includegraphics[width=\linewidth]{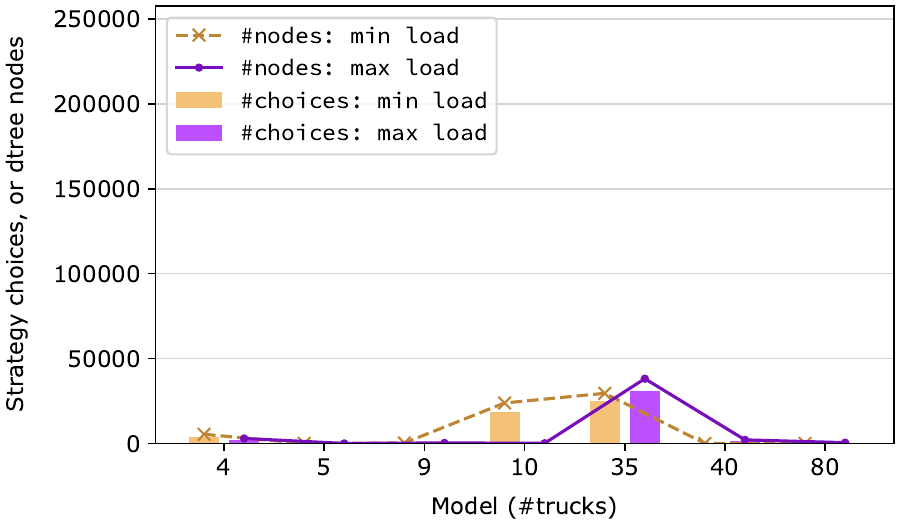}
    \end{subfigure}
    \vspace{1ex}

    \textsmaller[2]{Q-learning: 3M episodes}\\[.5ex]
    \begin{subfigure}{\WIDTH}
        \centering
        \smaller
        \includegraphics[width=\linewidth]{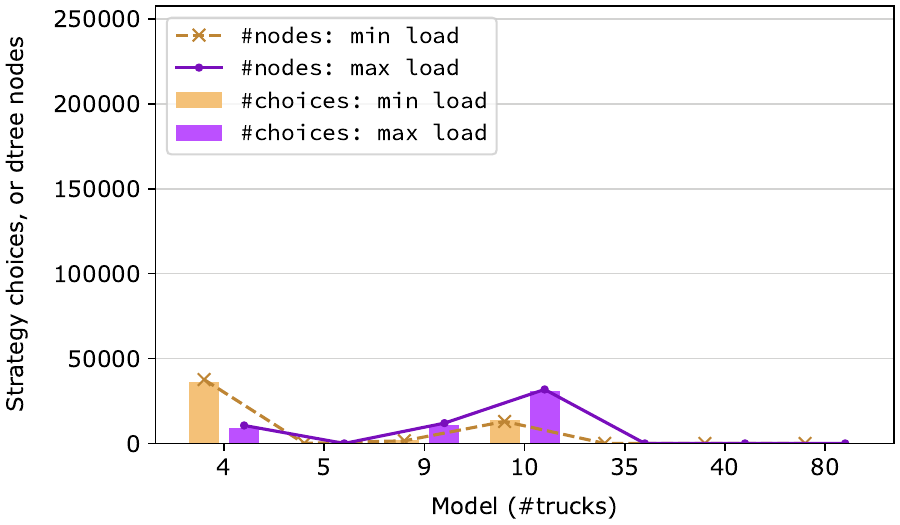}
    \end{subfigure}
    ~~
    \begin{subfigure}{\WIDTH}
        \centering
        \includegraphics[width=\linewidth]{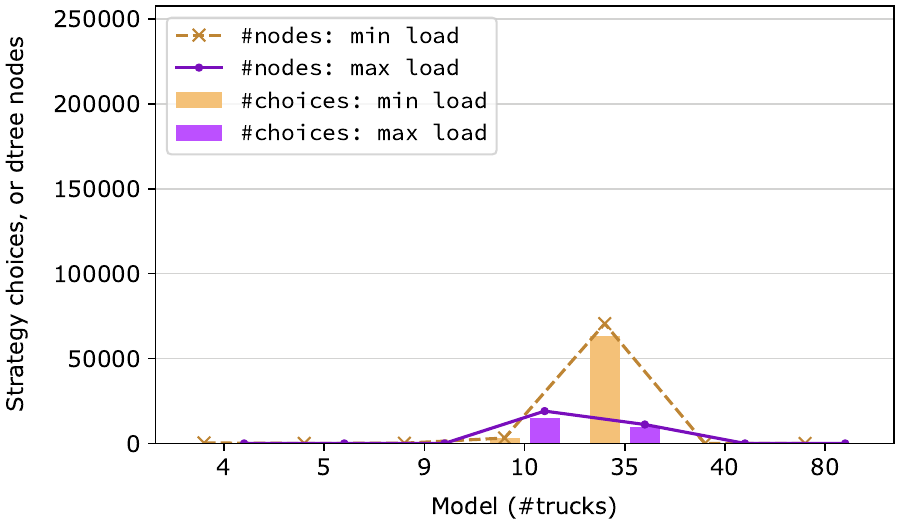}
    \end{subfigure}
    \vspace{1ex}


    \caption{Number of strategy choices vs.\ decision tree nodes}
    \label{fig:strategies_and_dtrees}
\end{figure}

\begin{figure}[t]
    \centering
    \def\WIDTH{.47\linewidth}
    
    \begin{subfigure}[t]{\WIDTH}
        \centering
        \smaller[2]
        \caption{Model 4, max load, LSS\,10k\,1k, \PO}
        \label{fig:DTs:4_LSS_PO}
        \includegraphics[width=.9\linewidth,trim={22pt 22pt 32pt 20pt},clip]{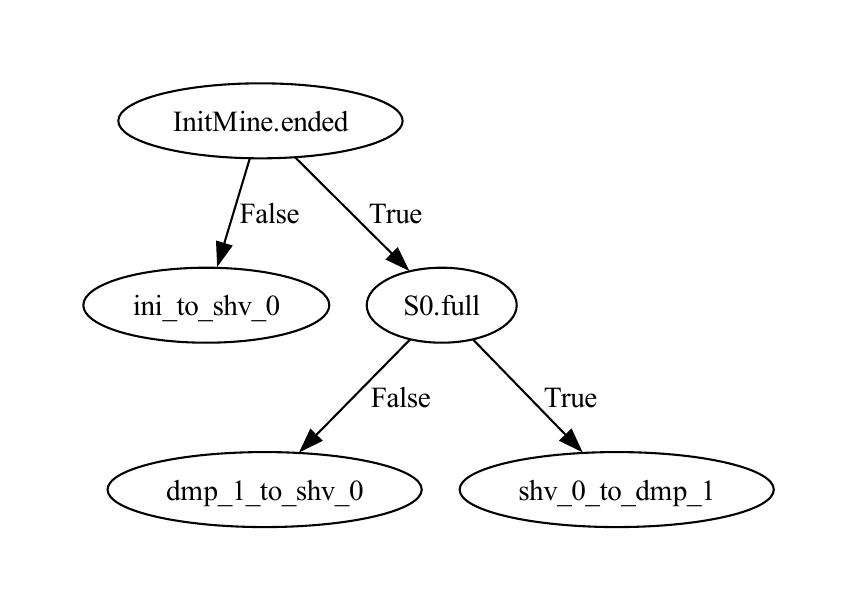}
    \end{subfigure}
    ~~
    \begin{subfigure}[t]{\WIDTH}
        \centering
        \smaller
        \caption{Model 5, max load, LSS\,10k\,1k, \FO}
        \label{fig:DTs:5_LSS_FO}
        \vspace{2ex}
        \includegraphics[width=\linewidth]{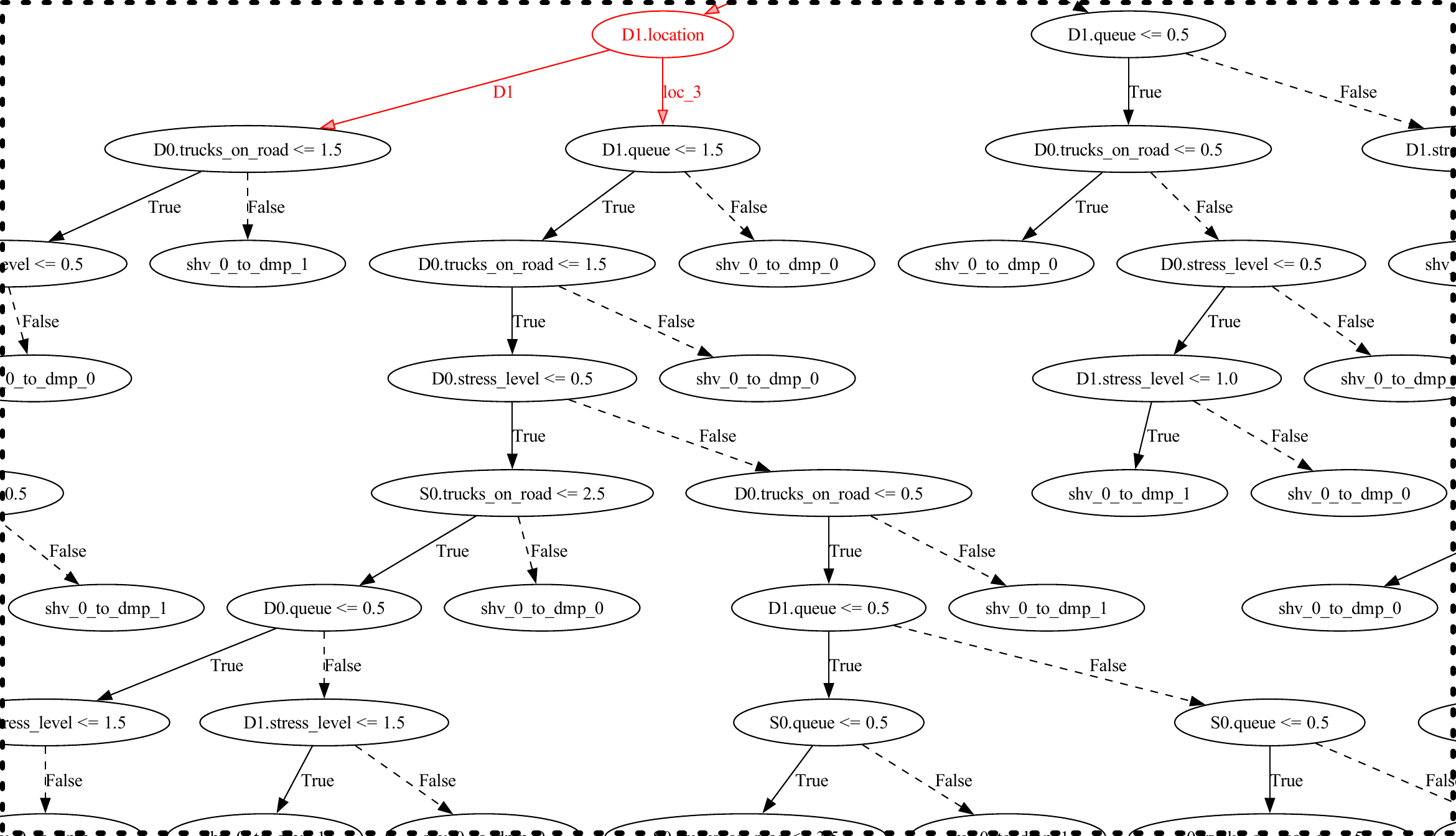}
    \end{subfigure}

    \caption{Two decision trees of LSS\,10k\,1k} 
    \label{fig:DTs}
\end{figure}

\paragraph{Results for strategies.}
We also study the strategies---and corresponding decision trees---that \modes synthesised to achieve the loads shown in \cref{fig:2D_plots,fig:3D_plots}.
For each model, partial- or fully-observable state, and minimum- or maximum-load objective, \Cref{fig:strategies_and_dtrees} compares the number of choices in a strategy vs.\ the number of nodes of the decision tree built for it.
We observe that decision trees achieve a slight compression (i.e.\ they contain fewer nodes than there are choices of the corresponding strategy) only in the case of \FO models.
This was expected, since fully-observable models contain variables that may be of no use to (minimise or) maximise the load, so removing them from the picture increases the entropy---we provide a concrete example in \Cref{fig:DTs:5_LSS_FO}.

Moreover, the strategies found by Q-learning contain much fewer choices than those of LSS (but note that for models $\geqslant35$ with \FO states, and $\geqslant40$ with \PO states, Q-learning failed to produce any strategy).
In several cases when computing minimum loads, Q-learning failed to learn non-zero values for any relevant states, resulting in empty strategies (since the choices had to be made randomly for lack of information).
As already discussed, the minimum and maximum loads achieved by this algorithm in its two configurations are quite close to the uniform random strategy.
And while Q-learning with a 3M-episodes budget did achieve a higher maximum load than the uniform, this is still comparable to the loads achieved by LSS.
Since, on top of this, Q-learning consumes much more memory and runtime than both LSS and the uniform random strategy, the fact that it produces smaller decision trees is arguably not a useful advantage here.

Finally, \Cref{fig:DTs} shows decision trees built by \dtcontrol from strategies synthesised by \modes running LSS\,10k\,1k on models 4 and 5 for maximum load (results for minimum are similar).
The DT from \Cref{fig:DTs:4_LSS_PO} is small enough to follow its logic.
Its first action is to initialise the mine at the (ore) shovel 0.
Then, an action is chosen depending on the state of that shovel:
if it is full, send a truck to transport useful material to the (ore) dump 1;
otherwise, send a truck to shovel 0, awaiting for it to fill up.
This simplistic decision tree reflects the six strategy choices that \modes found during SMC following LSS\,10k\,1k on \PO model 4.

The DT of \Cref{fig:DTs:4_LSS_PO} is small because it was built from a \PO state, and despite the fact that model 4 is of middle size: by its number of ore shovels and dumps it is effectively larger than models 5 and 9.
In contrast, model 5 is the smallest with 1 shovel and 2 dumps, none of which are for ore.
Notwithstanding, the snippet shown in \Cref{fig:DTs:5_LSS_FO}---built from a \FO state---comes from a DT containing 269 nodes and 268 decisions.
This includes the values of variables such as full shovels as discussed above, but also the number of trucks en route, queued at a shovel, or at a dump, and even the ``program counter'' of the \lstinline|DumpSystem| process that should actually be irrelevant (marked in red in \Cref{fig:DTs:5_LSS_FO}).
Compare this to the DT that both LSS\,10k\,1k and LSS\,100k\,10k build for maximum load using \PO states, and which consists of three nodes: do either \scalebox{.9}{\ttfamily ini\_to\_dmp\_0} (initialise the system) or else \scalebox{.9}{\ttfamily shv\_0\_to\_dmp\_0}.
This strategy sends a truck from \emph{the} shovel to its closest dump, thus maximising the load in this toy example, which however becomes complicated if all model variables are observable as in the \FO case.
For min load and \PO state, LSS\,10k\,1k and LSS\,100k\,10k never choose to send to \scalebox{.9}{\ttfamily dmp\_0} but do it to \scalebox{.9}{\ttfamily dmp\_1} instead.

\paragraph{Effectiveness of LSS and QL.}
Since no model checker implements the analysis of time-bounded expected accumulated rewards on MA, we have no ``ground truth'' to judge the effectiveness of LSS and QL in our new implementation.
In order to have an idea of the possible performance of a model checker, we instead ran \mcsta for a rather similar property, namely the
expected accumulated \emph{reachability} reward where the goal is given by
the expiration of a randomly set timer.
The outcome was that \mcsta could check the three smallest models (not without some effort in the 4 trucks model for which it took over 15 minutes).
For all of the larger models (namely, models 10, 35, 40 and 80), it ran out of memory.

\section{Conclusion}

Motivated by a novel case study challenge to optimise operations in an open-pit mine, we investigated the state of the art in SMC-based optimisation methods available in tools from the probabilistic verification community.
We explained the LSS and QL approaches, with detailed pseudocode documenting our adaptations to the model of MA.
To improve the effectiveness of the analysis, and to obtain explainable results, we extended the \modes statistical model checker with support for partial observability to sample/learn based on selected model \emph{features}, with a memory-efficient approach to collecting strategy decisions---the first approach to turn LSS's strategy identifiers into useful information---and with a connection to \dtcontrol to obtain decision tree representations of complex strategies.
Based on a \modest model of the open-pit mining operations, we compared the effectiveness and performance of LSS and QL, with and without selecting model features.
We find that, somewhat surprisingly, the uniform random strategy is hard to beat by sampling or learning---but the methods are effective as evidenced by their ability to find strategies that succeed at \emph{minimising} the work done in the mine.
Identifying model features pays off, but always requires care and a good understanding of the case study at hand.
In comparing LSS and QL, we see that the former works better despite its more simplistic, uninformed approach.
In particular, LSS preserves the constant-memory property of SMC, while QL runs into the state space explosion problem just like standard model checking.
We conjecture that this effect is evidence that, for our case study, no small ``core''~\cite{KM20} exists that suffices to obtain good strategies while disregarding a large amount of the mine's behaviour.
Our results reinforce earlier data~\cite{HK22} hinting at learning-based approaches being inferior to LSS on a level playing field, but potentially being able to provide much better results when carefully tweaked, such as by employing neural networks instead of explicit Q-tables and initialising episodes from non-initial states if the model allows to do so~\cite{GGHHKMMSW23}.
This motivates us to compare with neural network-based methods in future work.
Successful results of the partition-refinement learning method implemented in \tool{Uppaal Stratego}~\cite{JLM22} suggest another approach worth comparing~to.

\begin{credits}
  \paragraph{Data availability.}
  The models and tools/scripts to reproduce our experimental evaluation are archived and available at DOI \href{https://doi.org/10.5281/zenodo.13327230}{10.5281/zenodo.13327230}~\cite{PaperArtifact}.

  \paragraph{\ackname}
  We are grateful to Mat\'ias D.\ Lee and Joaqu\'in Feltes for
  discussions and insights on early versions of the open-pit mine
  model.

  \paragraph{Funding.}
  This work was supported by Agencia I$+$D$+$i grant PICT
  2022-09-00580 ({\scriptsize CoSMoSS}), the European Union's Horizon 2020
  research and innovation programme under the Marie
  Sk{\l}odowska-Curie grant agreements 101008233 ({\scriptsize MISSION}) and 101067199 ({\scriptsize ProSVED}), the Interreg North Sea project STORM\_SAFE, the Next\-GenerationEU projects D53D23008400006 ({\scriptsize SMARTITUDE}) under MUR PRIN 2022 and PE00000014 ({\scriptsize SERICS}) under MUR PNRR,
  NWO VIDI grant VI.Vidi.223.110 ({\scriptsize TruSTy}),
  and SeCyT-UNC grant 33620230100384CB ({\scriptsize MECANO}).
\end{credits}

\bibliographystyle{splncs04}
\bibliography{paper}

\begin{thebibliography}{10}
\providecommand{\url}[1]{\texttt{#1}}
\providecommand{\urlprefix}{URL }
\providecommand{\doi}[1]{https://doi.org/#1}

\bibitem{AP18}
Agha, G., Palmskog, K.: A survey of statistical model checking. {ACM} Trans.
  Model. Comput. Simul.  \textbf{28}(1),  6:1--6:39 (2018).
  \doi{10.1145/3158668}

\bibitem{AG02}
Alarie, S., Gamache, M.: Overview of solution strategies used in truck
  dispatching systems for open pit mines. International Journal of Surface
  Mining, Reclamation and Environment  \textbf{16}(1),  59--76 (2002).
  \doi{10.1076/ijsm.16.1.59.3408}

\bibitem{AD94}
Alur, R., Dill, D.L.: A theory of timed automata. Theor. Comput. Sci.
  \textbf{126}(2),  183--235 (1994). \doi{10.1016/0304-3975(94)90010-8}

\bibitem{ABHK18}
Ashok, P., Butkova, Y., Hermanns, H., Kret{\'{\i}}nsk{\'{y}}, J.:
  Continuous-time {M}arkov decisions based on partial exploration. In: Lahiri,
  S.K., Wang, C. (eds.) 16th International Symposium on Automated Technology
  for Verification and Analysis ({ATVA}). Lecture Notes in Computer Science,
  vol. 11138, pp. 317--334. Springer (2018).
  \doi{10.1007/978-3-030-01090-4\_19}

\bibitem{AJKWWY21}
Ashok, P., Jackermeier, M., Kret{\'{\i}}nsk{\'{y}}, J., Weinhuber, C.,
  Weininger, M., Yadav, M.: {dtControl} 2.0: Explainable strategy
  representation via decision tree learning steered by experts. In: Groote,
  J.F., Larsen, K.G. (eds.) 27th International Conference on Tools and
  Algorithms for the Construction and Analysis of Systems ({TACAS}). Lecture
  Notes in Computer Science, vol. 12652, pp. 326--345. Springer (2021).
  \doi{10.1007/978-3-030-72013-1\_17}

\bibitem{BAFK18}
Baier, C., de~Alfaro, L., Forejt, V., Kwiatkowska, M.: Model checking
  probabilistic systems. In: Clarke, E.M., Henzinger, T.A., Veith, H., Bloem,
  R. (eds.) Handbook of Model Checking, pp. 963--999. Springer (2018).
  \doi{10.1007/978-3-319-10575-8\_28}

\bibitem{BDL04}
Behrmann, G., David, A., Larsen, K.G.: A tutorial on {U}ppaal. In: Bernardo,
  M., Corradini, F. (eds.) International School on Formal Methods for the
  Design of Computer, Communication and Software Systems ({SFM-RT}). Lecture
  Notes in Computer Science, vol.~3185, pp. 200--236. Springer (2004).
  \doi{10.1007/978-3-540-30080-9\_7}

\bibitem{Bel57}
Bellman, R.: A {M}arkovian decision process. Journal of Mathematics and
  Mechanics  \textbf{6}(5),  679--684 (1957)

\bibitem{BDHK06}
Bohnenkamp, H.C., D'Argenio, P.R., Hermanns, H., Katoen, J.P.: {MoDeST}: A
  compositional modeling formalism for hard and softly timed systems. {IEEE}
  Trans. Software Eng.  \textbf{32}(10),  812--830 (2006).
  \doi{10.1109/TSE.2006.104}

\bibitem{PaperArtifact}
Budde, C.E., D'Argenio, P.R., Hartmanns, A.: Artifact for digging for decision
  trees: A case study in strategy sampling and learning. Zenodo (2024).
  \doi{10.5281/zenodo.13327230}

\bibitem{BDHS20}
Budde, C.E., D'Argenio, P.R., Hartmanns, A., Sedwards, S.: An efficient
  statistical model checker for nondeterminism and rare events. Int. J. Softw.
  Tools Technol. Transf.  \textbf{22}(6),  759--780 (2020).
  \doi{10.1007/S10009-020-00563-2}

\bibitem{BDLMPLW12}
Bulychev, P.E., David, A., Larsen, K.G., Mikucionis, M., Poulsen, D.B., Legay,
  A., Wang, Z.: {U}ppaal-{SMC}: Statistical model checking for priced timed
  automata. In: Wiklicky, H., Massink, M. (eds.) 10th Workshop on Quantitative
  Aspects of Programming Languages and Systems ({QAPL}). {EPTCS}, vol.~85, pp.
  1--16 (2012). \doi{10.4204/EPTCS.85.1}

\bibitem{BF19}
Butkova, Y., Fox, G.: Optimal time-bounded reachability analysis for concurrent
  systems. In: Vojnar, T., Zhang, L. (eds.) 25th International Conference on
  Tools and Algorithms for the Construction and Analysis of Systems ({TACAS}).
  Lecture Notes in Computer Science, vol. 11428, pp. 191--208. Springer (2019).
  \doi{10.1007/978-3-030-17465-1\_11}

\bibitem{BHH21}
Butkova, Y., Hartmanns, A., Hermanns, H.: A {M}odest approach to {M}arkov
  automata. {ACM} Trans. Model. Comput. Simul.  \textbf{31}(3),  14:1--14:34
  (2021). \doi{10.1145/3449355}

\bibitem{BHHK15}
Butkova, Y., Hatefi, H., Hermanns, H., Krc{\'{a}}l, J.: Optimal continuous time
  {M}arkov decisions. In: Finkbeiner, B., Pu, G., Zhang, L. (eds.) 13th
  International Symposium on Automated Technology for Verification and Analysis
  ({ATVA}). Lecture Notes in Computer Science, vol.~9364, pp. 166--182.
  Springer (2015). \doi{10.1007/978-3-319-24953-7\_12}

\bibitem{BR11}
Bäuerle, N., Rieder, U.: {M}arkov Decision Processes with Applications to
  Finance. Springer (2011). \doi{10.1007/978-3-642-18324-9}

\bibitem{DHS18}
D'Argenio, P.R., Hartmanns, A., Sedwards, S.: Lightweight statistical model
  checking in nondeterministic continuous time. In: Margaria, T., Steffen, B.
  (eds.) 8th International Symposium on Leveraging Applications of Formal
  Methods, Verification and Validation ({ISoLA}). Lecture Notes in Computer
  Science, vol. 11245, pp. 336--353. Springer (2018).
  \doi{10.1007/978-3-030-03421-4\_22}

\bibitem{DLST15}
D'Argenio, P.R., Legay, A., Sedwards, S., Traonouez, L.M.: Smart sampling for
  lightweight verification of {M}arkov decision processes. Int. J. Softw. Tools
  Technol. Transf.  \textbf{17}(4),  469--484 (2015).
  \doi{10.1007/S10009-015-0383-0}

\bibitem{DJLMT15}
David, A., Jensen, P.G., Larsen, K.G., Mikucionis, M., Taankvist, J.H.:
  {U}ppaal {S}tratego. In: Baier, C., Tinelli, C. (eds.) 21st International
  Conference on Tools and Algorithms for the Construction and Analysis of
  Systems ({TACAS}). Lecture Notes in Computer Science, vol.~9035, pp.
  206--211. Springer (2015). \doi{10.1007/978-3-662-46681-0\_16}

\bibitem{EHZ10}
Eisentraut, C., Hermanns, H., Zhang, L.: On probabilistic automata in
  continuous time. In: 25th Annual {IEEE} Symposium on Logic in Computer
  Science ({LICS}). pp. 342--351. {IEEE} Computer Society (2010).
  \doi{10.1109/LICS.2010.41}

\bibitem{GGHHKMMSW23}
Gros, T.P., Gro{\ss}, J., H{\"{o}}ller, D., Hoffmann, J., Klauck, M., Meerkamp,
  H., M{\"{u}}ller, N.J., Schaller, L., Wolf, V.: {DSMC} evaluation stages:
  Fostering robust and safe behavior in deep reinforcement learning -- extended
  version. {ACM} Trans. Model. Comput. Simul.  \textbf{33}(4),  17:1--17:28
  (2023). \doi{10.1145/3607198}

\bibitem{HHHK13}
Hahn, E.M., Hartmanns, A., Hermanns, H., Katoen, J.P.: A compositional
  modelling and analysis framework for stochastic hybrid systems. Formal
  Methods Syst. Des.  \textbf{43}(2),  191--232 (2013).
  \doi{10.1007/S10703-012-0167-Z}

\bibitem{HH14}
Hartmanns, A., Hermanns, H.: The {M}odest {T}oolset: An integrated environment
  for quantitative modelling and verification. In: {\'{A}}brah{\'{a}}m, E.,
  Havelund, K. (eds.) 20th International Conference on Tools and Algorithms for
  the Construction and Analysis of Systems ({TACAS}). Lecture Notes in Computer
  Science, vol.~8413, pp. 593--598. Springer (2014).
  \doi{10.1007/978-3-642-54862-8\_51}

\bibitem{HH19}
Hartmanns, A., Hermanns, H.: A {M}odest {M}arkov automata tutorial. In:
  Kr{\"{o}}tzsch, M., Stepanova, D. (eds.) 15th International Reasoning Web
  Summer School on Explainable Artificial Intelligence ({RW}). Lecture Notes in
  Computer Science, vol. 11810, pp. 250--276. Springer (2019).
  \doi{10.1007/978-3-030-31423-1\_8}

\bibitem{HJQW23}
Hartmanns, A., Junges, S., Quatmann, T., Weininger, M.: A practitioner's guide
  to {MDP} model checking algorithms. In: Sankaranarayanan, S., Sharygina, N.
  (eds.) 29th International Conference on Tools and Algorithms for the
  Construction and Analysis of Systems ({TACAS}). Lecture Notes in Computer
  Science, vol. 13993, pp. 469--488. Springer (2023).
  \doi{10.1007/978-3-031-30823-9\_24}

\bibitem{HK22}
Hartmanns, A., Klauck, M.: The {M}odest state of learning, sampling, and
  verifying strategies. In: Margaria, T., Steffen, B. (eds.) 11th International
  Symposium on Leveraging Applications of Formal Methods, Verification and
  Validation ({ISoLA}). Lecture Notes in Computer Science, vol. 13703, pp.
  406--432. Springer (2022). \doi{10.1007/978-3-031-19759-8\_25}

\bibitem{Hat17}
Hatefi-Ardakani, H.: Finite horizon analysis of {M}arkov automata. Ph.D.
  thesis, Saarland University, Germany (2017),
  \url{http://scidok.sulb.uni-saarland.de/volltexte/2017/6743/}

\bibitem{HJKQV22}
Hensel, C., Junges, S., Katoen, J.P., Quatmann, T., Volk, M.: The probabilistic
  model checker {S}torm. Int. J. Softw. Tools Technol. Transf.  \textbf{24}(4),
   589--610 (2022). \doi{10.1007/S10009-021-00633-Z}

\bibitem{JLM22}
Jensen, P.G., Larsen, K.G., Mikucionis, M.: Playing wordle with uppaal
  stratego. In: Jansen, N., Stoelinga, M., van~den Bos, P. (eds.) A Journey
  from Process Algebra via Timed Automata to Model Learning - Essays Dedicated
  to Frits Vaandrager on the Occasion of His 60th Birthday. Lecture Notes in
  Computer Science, vol. 13560, pp. 283--305. Springer (2022).
  \doi{10.1007/978-3-031-15629-8\_15}

\bibitem{KLC98}
Kaelbling, L.P., Littman, M.L., Cassandra, A.R.: Planning and acting in
  partially observable stochastic domains. Artif. Intell.  \textbf{101}(1-2),
  99--134 (1998). \doi{10.1016/S0004-3702(98)00023-X}

\bibitem{KM20}
Kret{\'{\i}}nsk{\'{y}}, J., Meggendorfer, T.: Of cores: A partial-exploration
  framework for {M}arkov decision processes. Log. Methods Comput. Sci.
  \textbf{16}(4) (2020), \url{https://lmcs.episciences.org/6833}

\bibitem{KNP11}
Kwiatkowska, M.Z., Norman, G., Parker, D.: {PRISM} 4.0: Verification of
  probabilistic real-time systems. In: Gopalakrishnan, G., Qadeer, S. (eds.)
  23rd International Conference on Computer Aided Verification ({CAV}). Lecture
  Notes in Computer Science, vol.~6806, pp. 585--591. Springer (2011).
  \doi{10.1007/978-3-642-22110-1\_47}

\bibitem{Law15}
Law, A.M.: Simulation Modeling and Analysis. McGraw-Hill series in industrial
  engineering and management science, McGraw-Hill Education, 5th edn. (2015)

\bibitem{LST14}
Legay, A., Sedwards, S., Traonouez, L.M.: Scalable verification of {M}arkov
  decision processes. In: Canal, C., Idani, A. (eds.) Software Engineering and
  Formal Methods -- Revised Selected Papers of the {SEFM} 2014 Collocated
  Workshops: {HOFM}, {SAFOME}, {OpenCert}, {MoKMaSD}, and {WS-FMDS}. Lecture
  Notes in Computer Science, vol.~8938, pp. 350--362. Springer (2014).
  \doi{10.1007/978-3-319-15201-1\_23}

\bibitem{MA17}
Moradi~Afrapoli, A., Askari-Nasab, H.: Mining fleet management systems: a
  review of models and algorithms. International Journal of Mining, Reclamation
  and Environment  \textbf{33}(1),  42--60 (2019).
  \doi{10.1080/17480930.2017.1336607}

\bibitem{NPZ17}
Norman, G., Parker, D., Zou, X.: Verification and control of partially
  observable probabilistic systems. Real Time Syst.  \textbf{53}(3),  354--402
  (2017). \doi{10.1007/S11241-017-9269-4}

\bibitem{Put94}
Puterman, M.L.: {M}arkov Decision Processes: Discrete Stochastic Dynamic
  Programming. Wiley Series in Probability and Statistics, Wiley (1994).
  \doi{10.1002/9780470316887}

\bibitem{SB98}
Sutton, R.S., Barto, A.G.: Reinforcement learning -- An introduction. Adaptive
  computation and machine learning, {MIT} Press (1998)

\bibitem{WD92}
Watkins, C.J.C.H., Dayan, P.: Q-learning. Mach. Learn.  \textbf{8},  279--292
  (1992). \doi{10.1007/BF00992698}

\end{thebibliography}

\end{document}